\documentclass[epj,final]{svjour}
\usepackage{graphics}

\begin{document}
\title{Extensive Simulations for Longest Common Subsequences}
\subtitle{Finite Size Scaling, a Cavity Solution, and 
Configuration Space properties}

\author{J.~Boutet de Monvel}

\institute{Forschungszentrum BiBos, Fakult\H{a}t f\H{u}r Physik\\
Universit\H{a}t Bielefeld, 33615 Bielefeld, Germany\\
\email{boutet@physik.uni-bielefeld.de}}

\offprints{J. Boutet de Monvel}

\date{\today \\{\it Submitted to EPJB}}

\abstract{Given two strings X and Y of $N$ and $M$ characters respectively, 
the Longest Common Subsequence (LCS) Problem asks for the longest sequence 
of (non-contiguous) matches between X and Y.
Using extensive Monte Carlo simulations for this problem,
we find a finite size scaling law of the form
$E(L_N)/N =\gamma_S + A_S/(\ln{N\sqrt{N}})+...$ for the average LCS length
of two random strings of size $N$ over $S$ letters. We provide precise 
estimates of $\gamma_S$ for $2\le S\le 15$.
We consider also a related Bernoulli Matching model where the different entries 
of an $N\times M$ array are occupied with a match {\it independently} with 
probability $1/S$. On the basis of a cavity-like analysis we find that the 
length of a longest sequence of matches in that case behaves as 
$L_{NM}^B\sim \gamma^B_S(r) N$ where $r=M/N$ and
$\gamma^B_S(r)=(2\sqrt{rS}-r-1)/(S-1)$. 
This formula agrees very well with our numerical computations. It provides 
a very good approximation for the Random String model, the approximation 
getting more accurate as $S$ increases. The question of the ``universality class'' of the LCS problem is also considered. Our results for the Bernoulli 
Matching model show very good agreement with the scaling predictions of \cite{HwaLassig96_PRL} for Needleman-Wunsch sequence alignment.
We find however that the variance of the LCS length has a scaling different 
from $Var(L_N)\approx N^{2/3}$ in the Random String model, suggesting 
that long-ranged  correlations among the matches are relevant in 
this model.
We finally study the ``ground state'' properties of this problem. We find
that the number ${\cal N}_{LCS}$ of solutions typically grows
exponentially with $N$. In other words, this system does not 
satisfy ``Nernst's principle''. This is also reflected at the level of 
the overlap between two LCSs chosen at random, which is found to be 
self averaging and to aproach a definite value $q_S<1$ as $N\to \infty$.   
}

\maketitle
\section{Introduction}
\label{intro}

Let $X=(X_1,...,X_N)$ and $Y=(Y_1,...,Y_M)$ be two strings of characters.
Here the $X_i$'s and $Y_j$'s are letters of a given alphabet, which will 
be assumed throughout this paper to be finite and of fixed size $S\ge 2$.
The Longest Common Subsequence problem, which we shall refer to as 
the LCS problem, consists of finding a sequence of letters which appears as a
subsequence of both $X$ and $Y$, and which is of maximal size. Equivalently 
one can ask for two sequences $1\le i_1<...<i_L\le N$ and 
$1\le j_1<...<j_L\le M$ such that $X_{i_k}=Y_{j_k}$, $1\le k \le L$ and $L$ 
is maximal.

The length of a LCS can be viewed as a natural measure of the ``proximity'' 
of different strings of letters. It is an example of the
``best sequence alignments'' which are of use in biology, 
in tests for comparing long molecules such as proteins and nucleic 
acids \cite{NeedlemanWunsch70_JMB,Sankoff&Al76_JME,Waterman94_PTRSLB}. 

It is also an important problem in computer science, as the length of a
LCS of two strings is closely related to the number of editing operations 
(insertions/deletions) which are necessary to transform one string into the other
(the so called ``string-edit'' distance) \cite{WagnerFisher74_JACM}. A large
number of variants and applications of the LCS problem are also described in 
\cite{SankoffKruskal83_Book}.

Another, less obvious motivation for the study of this problem comes  
from the fact that it can be formulated as a model of directed 
passage time percolation on a two dimensional (triangular) lattice 
\cite{Ukkonen85_IC,Alexander94_AAP}. 
To see this, consider the directed lattice whose vertices are the 
integer points $(ij)$, $0\le i\le N, 0\le j\le M$ and whose edges are 
the bonds formed by nearest neighbors together with the bonds of 
the form $\{(i-1,j-1),(ij)\}$, $1\le i \le N, 1\le j\le M$,
all of these bonds being oriented according to the positive direction 
of the axes. 
To each bond between nearest neighbors attach the weight $0$, 
and to each bond $\{(i-1,j-1),(ij)\}$ attach the weight $\delta_{X_i,Y_j}$, 
that is $1$ if $X_i=Y_j$, and to $0$ otherwise. Define the weight of 
any path on this lattice to be the sum of the bonds' weights along the path.
Then clearly a LCS between $X$ and $Y$ may be constructed from any directed 
path of maximum weight joining the point $(0,0)$ to the point $(N,M)$. 
If we interpret the weight of a bond as a time required for the passage of 
that bond, we seek the maximum rather than the minimum passage 
time from $(0,0)$ to $(N,M)$, but this is of no significance here. 

This paper is concerned with the stochastic version of the LCS problem, 
where one is given very long strings the letters of which are chosen at 
random, independently and uniformly in a given alphabet of size $S$. 
This problem has retained much attention 
\cite{ChvatalSankoff75_JAP,Deken79_DM,Steele82_SIAMJAM} (see also 
\cite{DancikPaterson94_STACS} for a recent review). The main issue 
is to understand the large $N$ behaviour of the LCS length of the 
$N$ first letters of $X$ and $Y$. Let $L_N$ be this number. 
Observing that the sequence $(L_N)$ is superadditive 
($L_{N_1+N_2}\ge L_{N_1}+L_{N_2}$), and using the martingale difference method,
one can prove in an elegant way \cite{Steele97_Book} that with 
probability one (for infinite strings), $L_N$ is asymptotic from below 
to $\gamma_S N$, where $0<\gamma_S\le 1$ is a constant whose exact value is
unknown. 
It has also been proved \cite{Alexander94_AAP,Rhee95_AAP} that the rate of
convergence of the expected ratio $E(L_N)/N$ to $\gamma_S$ is at least as fast 
as $O(\sqrt{\ln{N}/N})$. 

In the passage time percolation picture the weights attached to 
the bonds are correlated random variables (for example the occupation 
numbers of the matches on the corners of any rectangle of the lattice 
are obviously correlated). We consider also a related model where each bond 
$\{(i-1,j-1),(ij)\}$ is given a weight $1$ (resp. $0$) 
{\it independently} of the others with probabitlity $1/S$ (resp. $1-1/S$).
We shall refer to this model as the Bernoulli Matching model, and denote
by $L^{B}_N$ the maximum weight of a directed lattice path joining 
$(0,0)$ to $(N,N)$ (equivalently $L^{B}_N$ is the maximum $L$ for which there 
are sequences $1\le i_1<...<i_L\le N$ and $1\le j_1<...<j_L\le N$ such that 
$(i_k,j_k)$ is a match, $1\le k \le L$). We let $\gamma^{B}_S$ be
the limit $\lim_{N\to \infty} L^{B}_N/N$, which is shown to exist a.e. in 
exactly the same way as for $\gamma_S$. Also we note that Alexander's rate
result \cite{Alexander94_AAP} applies to $E(L^B_N)$ as well.

Much effort have been made to get bounds on $\gamma_S$ 
\cite{Deken79_DM,Dancik94_Thesis}, but there are still non negligible gaps 
between the known upper and lower bounds \cite{DancikPaterson94_STACS}.
Estimations of $\gamma_S$ based on numerical simulation are also 
available \cite{DancikPaterson94_STACS,Alexander94_AAP,Waterman94_PTRSLB} 
but apparently no attempt has been made to determine numerically the finite 
size corrections to the linear scaling law $E(L_N)\sim \gamma_S N$.

This paper presents the results of extensive Monte Carlo simulations 
for the LCS problem, showing that the difference $\gamma_S N - E(L_N)$ 
has a well defined asymptotic behaviour, allowing one to get 
precise estimates of $\gamma_S$ by extrapolation. 
The same finite size scaling law appears to hold for the Bernoulli Matching 
model, and we have obtained corresponding estimates for $\gamma^{B}_S$. 

We further considered the case where the strings $X$ and 
$Y$ are of different sizes, $N\ne M$. 
The relevant case occurs when $N$ and $M$ are large but comparable, 
namely $N,M \to \infty$, the ratio $r=M/N$ being fixed ($r>0$). 
Let $L_N(r)=L_{N,[rN]}$ be the length of a LCS of $X_1,...,X_N$ and 
$Y_1,...,Y_{[rN]}$. Then with probability $1$, one has\\ 
$\lim_{N\to \infty} L_N(r)/N=\gamma_S(r)$ where $0<\gamma_S(r)\le 1$.
Of course $\gamma_S(1)=\gamma_S$, and the function $\gamma_S(r)$ 
has the obvious symmetry property $\gamma_S(1/r)=1/r \gamma_S(r)$.
In the picture of directed percolation $r$ is given by  
$\tan(\pi/4+\phi)$ where $\phi\in [-\pi/4,\pi/4]$ is the angle between the 
direction of interest and the first bisector, and the object of interest 
is the set of points which are ``wet'' at time $t$, defined here 
to be the set $C_t=\{(ij): L_{i,j}\le t\}$. 
As $t\to \infty$ (for $N$ and $M$ infinite) 
the set $C_t/t$ is asymptotically delimited by the curve of polar 
equation $\rho(\phi)=\sqrt{1+r(\phi)^2}/\gamma_S(r(\phi))$.
The above symmetry property reflects the fact that $C_t$ is 
asymptotically symmetric with respect to the first bissector.
A percolation transition occurs in this problem when $r=r_c=S$, namely  
$\gamma_S(r)=1$ for $r\ge S$ while $\gamma_S(r)<1$ for $r<S$.
By symmetry we have another transition at $r=1/r_c=1/S$, such that
$\gamma_S(r)=r$ for $r\le 1/S$ and $\gamma_S(r)<r$ for $r>1/S$.
Analogous comments apply to the Bernoulli Matching model. 
In that case we provide a simple analytic expression for the 
corresponding function $\gamma^B_S(r)$, which is derived (see section
(\ref{N_ne_M_case}) below) on the basis of a cavity-like analysis 
of the LCS problem. 
The cavity method is an approximation scheme generally considered to be 
appropriate for describing the mean field theory of disordered systems 
(such as spin glasses) \cite{MezardParisiVirasoro87_Book}. 
The Bernoulli Matching model is not a mean field model however, but 
really a two dimensional percolation model, and by ``cavity'' we 
mean the following: First, the properties of the system can be 
computed by use of a recursion formula. 
This is equation (\ref{Lij_rec}) given below,
which is valid for the Random String model as well as 
for the Bernoulli Matching model. 
Second, a decorrelation, or ``clustering'' property 
\cite{MezardParisiVirasoro87_Book} happens to hold in the 
Bernoulli Matching model, allowing the recursion formula 
to be solved at large $N,M$ by use of a self-consistent approximation. 
This leads to an expression of $\gamma_S^B(r)$ in very good agreement 
with our numerical results. 

We finally investigated the ``configuration space'' properties of this 
problem, which are most easily accessible by constructing what we call 
the LCS graph of given strings $X$ and $Y$. This structure is defined 
in section (\ref{config_space_properties}). It can be computed in a very 
efficient way, and it gives a direct access to properties of the set 
of LCSs of $X$ and $Y$, enabling one to compute such quantities as:

i) The total number ${\cal N}_{LCS}$ of LCSs of $X$ and $Y$.

ii) The average overlap between two LCSs chosen at random among the set of 
LCSs of $X$ and $Y$.

iii) The distribution of the distance between two successive matches in 
a LCS. By distance we mean here the Manhattan distance
$|i_1-i_2|+|j_1-j_2|$ for given points $(i_1j_1),(i_2j_2)$.

iv) The mean square ``displacement'' with respect to the first
bissector, of the matches along a LCS, where 
(following \cite{HwaLassig96_PRL}) the displacement coordinate 
of a point $(ij)$ is defined to be $i-j$.

These type of computations are of interest because they provide 
informations on the structure of the set of solutions which in other 
systems may be very difficult to obtain. 
For example our computations show that typical random strings 
have {\it many} common subsequences of maximum length. Their number 
typically grows exponentially with $N$, i.e. the ground state entropy 
of this system is not zero. We provide estimates of this entropy
and of the typical overlap between two randomly chosen LCSs for 
several values of $S$. 
Properties such as iv) are of physical interest as they depend 
on long-ranged correlations among the matches in a LCS, and they 
characterize the ``universality class'' of the LCS problem. 
This question has been recently analysed by Hwa and L\H{a}ssig
\cite{HwaLassig96_PRL} who showed that the percolation formulation 
of the LCS problem (and more generally of Needleman-Wunsch sequence 
alignment described below) can be treated in the continuum limit as
a model of directed polymer in a quenched random medium. 
In this analogy each directed path on the above defined 
lattice is assigned an energy $-W$, where $W$ is the weight 
of the path. The statistics of these paths is taken to be the 
Boltzmann-Gibbs distribution \cite{ZhangMarr95_JTB}.
The LCS length then corresponds to the ground state energy 
of the ``bridge'' from $(0,0)$ to $(N,M)$. 
In the case of the Bernoulli Matching model, this leads to a 
complete characterization of the universality class of the model: The
continuum limit is described by the well studied 2D-directed path
(or equivalently the 1D-random walk) in a Gaussian random potential. 
The fluctuations of $L^B_N$ and of the ``displacement'' $i-j$ along
the optimal paths are governed by exactly known universal exponents 
$\omega=1/3$ and $\zeta=2/3$ respectively.
Hence $Var(L_N^B)$ should grow asymptotically as $N^{2/3}$ 
and the mean square displacement as $N^{4/3}$. 
Our numerical results agree very well with these predictions.
The question of the universality class of the Random String model is more
subtle, as this model involves long ranged correlations in the disorder.
Hwa and L\H{a}ssig provide evidence that these correlations 
are not relevant in the continuum limit for a range of the defining 
parameters of Needleman-Wunsch alignment. 
In the regime corresponding to the LCS problem (which was 
not considered in \cite{HwaLassig96_PRL}), our results only partly 
supports the above predictions: 
The measured mean square displacement for the Random String model 
show no deviation from the superdiffusive $N^{4/3}$ scaling.
The behaviour of $Var(L_N)$ is close to, but significantly different 
from $N^{2/3}$, suggesting that correlations among the matches in the 
Random String model {\it are} relevant to the universality class of 
the LCS problem. 
It should not be considered a surprise that the scaling relation 
$\omega=2\zeta -1$ appears invalidated by our results.
This scaling relation is known to be intimately connected
with Galilean invariance \cite{MedinaHwaKardar89_PRA}. 
In the formulation of the Random String LCS problem as a 1D-random 
walk, long-range {\it temporal} correlations are present in 
the random potential, and Galilean invariance is broken. 
What is surprising is that only the fluctuations of the ground 
state energy show a scaling affected by these correlations.
This is left to the reader as an interesting open question.

We close this introduction by explaining the position of the 
LCS problem with respect to sequence alignment methods in molecular biology. 
The purpose of these methods is to provide efficient tools for the 
detection of relevant similarities among DNA molecules or among proteins. 
Relevance refers here to finding the functional and evolutionary 
relationships between these molecules, and is a main biological issue.
This problem is the source of a rich interplay between biology and 
computational sciences (see \cite{Waterman89_Book} for reviews). 
Even if determining what is the ``best alignment'' of two
sequences for biological purposes remains in part a matter of art, 
standard comparison algorithms are widely used by biologists. 
These algorithms are very useful to confront a newly discovered 
DNA molecule or protein to the huge existing databases of known 
molecules (and then to infer the possible functional properties 
of the new molecule). 
The LCS problem corresponds to a class of alignment algorithms 
discovered by Needleman and Wunsch \cite{NeedlemanWunsch70_JMB},
which provided the first systematic tool for taking into 
account the insertions and deletions which naturally occurs in 
the evolution of biological sequences.
To describe this approach consider again the percolation formulation 
of the LCS problem. An alignment of the strings
$X$ and $Y$ is viewed as a directed path on the the lattice defined 
above, tracing a possible ``evolution'' from $X$ to $Y$: 
Each diagonal bond (ending at $(ij)$) on the path represents a substitution 
of the letter $Y_j$ to the letter $X_i$ (if $(ij)$ is a match $X_i$ is 
left unchanged). Horizontal and vertical bonds represents respectively 
deletions and insertions, also termed as indel operations, or ``gaps''. 
In this way each directed path from $(0,0)$ to $(N,M)$ corresponds to a well 
defined sequence of edit operations transforming $X$ into $Y$ (or equivalently 
$Y$ into $X$), which is the usual definition of an ``alignment''. 
A given path $\gamma$ is assigned a score $W(\gamma)$, which is 
defined (in the simplest version of the model) by weighting each 
substitution along $\gamma$ with a matching function 
$s(X_i,Y_j)$, and each gap with a penalty $-\delta (\delta>0)$. 
A common choice for $s(X_i,Y_j)$ is to assign 
a score $1$ to a match $X_i=Y_j$ and a penalty $-\mu$ ($\mu>0$) to 
a mismatch $X_i\ne Y_j$.
The optimal alignments are determined by maximisation of the score 
$W(\gamma)$. We are then facing a longest path problem very 
similar to the LCS problem. In particular the optimal score $W_{NM}$ 
from $(0,0)$ to $(N,M)$ can be computed in an efficient way using 
a straightforward adaptation of the dynamic programming algorithm 
of section (\ref{av_LCSlength}). 
Needleman-Wunsch sequence alignment is a {\it global} alignment method, 
since the whole strings $X$ and $Y$ are aligned together. 
The optimal alignments are invariant by multiplying the matching 
function and the gap penalty by any positive constant.  
Moreover the numbers $N_{+}$, $N_{-}$, and $N_{g}$ of matches, 
mismatches, and gaps respectively along any directed path 
from $(0,0)$ to $(N,M)$ are related by $2N_{+}+2N_{-}+N_{g}=N+M$. 
Hence with the above choices the number of independent parameters is 
reduced to one: It is equivalent to maximize 
$W(\gamma)=N_{+}-\mu N_{-}-\delta N_{g}$ or to maximize  
${\tilde W}(\gamma)=N_{+}-\epsilon N_{g}$, where 
$\epsilon=(\delta-\mu/2)/(1+\mu)$. As $N,M\to \infty$ the 
modified optimal score behaves as ${\tilde W}_{NM}\sim a(\epsilon,r) N$ 
($r=M/N$), where $a(\epsilon,r)$ is a monotonous decreasing 
(demonstrably continuous) function of $\epsilon$. For $\epsilon\le -1/2$ 
the problem is trivial and ${\tilde W}_{NM}=-(N+M)\epsilon$.
When $-1/2<\epsilon<0$, it is always advantageous to change a mismatch
for two gaps (an insertion followed by a deletion). We may then assume
$2N_{+}+N_{gaps}=N+M$ and the problem reduces to maximizing $N_{+}$, i.e. 
to the LCS problem. In this region $a(\epsilon,r)$ interpolates linearly
from its value at $\epsilon=-1/2$ to its value at $\epsilon=0$. 
The case $\epsilon=0$ corresponds exactly to the LCS problem: Mismatches 
and gaps are then equivalent as regards to the score. 
Since gaps and mismatches are known both to occur during evolution, 
and are not equivalent energetically, the biologically relevant 
region clearly lies within $\epsilon>0$. Hence the LCS problem represents
a natural (even if unrealistic) limit case of Needleman-Wunsch sequence 
alignment. 
It must be pointed out that for biological purposes 
(in particular for detecting weak similaritites between rather 
remote sequences), {\it local} rather than global alignment 
is often required. A powerful approach to local alignment is Smith-Waterman
algorithm \cite{SmithWaterman81_JMB}, which maximizes the score 
$W(\gamma)$ over {\it all} pairs of substrings (i.e. contiguous segments) 
of $X$ and $Y$. In the percolation picture, the end points of the paths
associated with local alignments are no longer fixed.
The gap and mismatch penalties are then really different
parameters and strongly influence the optimal alignments. 
In fact for random sequences Smith-Waterman alignment undergoes a 
phase transition from global to local alignment
\cite{ArratiaWaterman94_AAP,Waterman94_PTRSLB}:
For small $\delta$ and $\mu$, more precisely as long as $\delta$ 
and $\mu$ are such that the optimal score $W_{NM}$ obtained by global 
alignment is positive, we recover essentially Needleman-Wunsch alignment: 
For large $N,M$, the optimal Smith-Waterman score $H_{NM}$ satisfies 
$H_{NM}\approx W_{NM}$ with high probability. 
Note that the case $\delta=0$ reduces as before to the LCS problem.
For sufficiently high gap and mismatch penalties, global alignment 
leads to a negative score $W_{NM}$ growing linearly with $N,M$ in absolute 
value. A positive score can be achieved only by  small paths taking 
advantage of the local fluctuations in the density of matches. This is 
the genuinely local phase, where $H_{NM}$ grows only logarithmically 
with $N,M$. Clearly the LCS problem is no more relevant to this phase.
For example the exponential proliferation of solutions occuring in the 
LCS problem, relevant to the global phase, is  replaced
in the local phase by a small number of well characterized optimal 
and suboptimal alignments \cite{VingronWaterman94_JMB}.
The transition line between the global and the local phases, 
which separates the regions of positive and negative linear growth 
of the global score $W_{NM}$, is easily determined from the knowledge of 
$a(\epsilon,r)$ defined above. Interestingly, the neighborhood 
of this transition line is found empirically to be a most relevant 
$(\delta,\mu)$-region for biological purposes 
\cite{VingronWaterman94_JMB}. 
Hence the value $a(0,r)=\gamma_S(r)$ provides some valuable information 
thanks to the monotonicity of $a(\epsilon,r)$. 
More importantly, even if the biological relevance of purely global 
alignments is for the present difficult to address, clearly it is 
of interest to understand their statistical properties 
\cite{DrasdoHwaLassig98_condmat}. 
As the LCS problem corresponds in some sense to the ``most'' global 
case of sequence alignment, it deserves particular attention.

\section{The average length of a Longest Common Subsequence.}
\label{av_LCSlength}

There are several algorithms for computing the LCS length of two  
strings $X=(X_1,...,X_N)$ and $Y=(Y_1,...,Y_M)$. The best known 
is based on a dynamic programing approach as follows.
For $i,j\ge 1$, let $L_{ij}$ be the length of a LCS of 
$(X_1,...,X_i)$ and $(Y_1,...,Y_j)$. We call the matrix $(L_{ij})$
the LCS matrix of the given instance.
The strategy consists of using the fact that $L_{ij}$ can be readily 
computed if $L_{i-1,j-1},L_{i-1,j}$ and $L_{i,j-1}$ are known. 
Indeed one has $L_{ij}=L_{i-1,j-1}+1$ when $X_i=Y_j$, and
$L_{ij}=\max(L_{i-1,j},L_{i,j-1})$ when $X_i\ne Y_j$. In short
\begin{equation} \label{Lij_rec}
L_{ij}=\max(L_{i-1,j},L_{i,j-1},L_{i-1,j-1}+\delta_{X_i,Y_j}).
\end{equation}  
This recurrence relation, with the obvious initial conditions
$L_{i,0}=L_{0,j}=0$, provide a very simple and efficient 
way to compute the LCS matrix of $X$ and $Y$.
This algorithm arises also naturally in the passage 
time percolation picture. 
Indeed the LCS problem, viewed as a longest 
directed path problem as described above, has a natural formulation as
a linear programming problem. Relation (\ref{Lij_rec}) is nothing but the
solution to the dual program, which is 
\begin{equation}
\min L_{NN}-L_{00}
\end{equation}
for given numbers $L_{ij}, 0\le i,j\le N$ subject to the 
constraints  
\begin{eqnarray}
L_{ij}\ge L_{i-1,j-1}+\delta_{X_i,Y_j}, L_{ij}\ge L_{i-1,j}, 
L_{ij} \ge L_{i,j-1}, \nonumber \\ 
\indent 1\le i,j\le N,
\end{eqnarray}
and $L_{i,0}=L_{0,j}=0$.

The time required to compute the LCS matrix of $X$ and $Y$ using  
(\ref{Lij_rec}) is given essentially by the product $NM$. 
Of course the whole LCS matrix contains more information than needed 
to construct a LCS of $X$ and $Y$ or to compute their LCS length. 
More involved algorithms focuss attention on subsets of 
the {\it set of matches} of $X$ and $Y$, i.e. the set of points
$(ij)$ such that $X_i=Y_j$. These algorithms may achieve much better 
time bounds in some special cases. However no algorithm is known for 
the LCS problem which achieve a significantly better time bound than $O(NM)$ 
in the general case, or even in average when $X$ and $Y$ 
are two random strings over an alphabet of size $S\ge 2$. 
The fastest known algorithm is described in \cite{MasekPaterson80_JCSS}. 

Moreover relation (\ref{Lij_rec}) is highly suited
for a finite size scaling analysis of the LCS length, 
as it may be easily implemented in order to compute 
in time $O(N^2)$ and space $O(N)$ the {\it whole profile} of 
values $L_i=L_{i,i}, 1\le i\le N$ for any given instance.
Indeed, to compute the $i^{th}$ line of the LCS matrix 
it is not necessary to have stored all the previous lines, 
since only the $(i-1)^{th}$ line is needed. 
This property also makes the computation of the LCS matrix 
parallelisable to some extent, and a significant speed up is
obtained in the case of very long strings by implementing 
(\ref{Lij_rec}) on a parallel machine.

\subsection{Finite size behaviour of $E(L_N)$.}

In order to measure the finite size behaviour of the average 
LCS length, we made a direct Monte Carlo evaluation of $E(L_N)$ 
for all $N$ up to a certain number and over large samples of random 
strings. Namely we computed averages of $L_N$ over $10^5$ instances 
for $N\le 1500$, and over $10^4$ instances for $1500\le N\le 10^4$.   
We then extrapolated these estimates to the large $N$ limit by using 
a $\chi^2$ analysis. In order to check the extrapolation procedure 
we performed a second series of experiments on a parallel computer,
over smaller samples of $30$ to $50$ instances, but for problem 
sizes up to $N=10^5$. 

We found that a very reliable extrapolation to the large $N$ limit
is obtained if one assumes a finite size behaviour of the form 
\begin{equation}\label{fsize_EL}
{E(L_N)\over N} = \gamma_S + {A_S\over \ln{N}\sqrt{N}}+\epsilon_N.
\end{equation}
Here $A_N$ is a negative constant and $\epsilon_N$ represents 
further corrections which we expect to be at most $O(1/N)$. 
To extract the precise asymptotic behaviour (if any) of 
$\epsilon_N$ would certainly require improvement on the 
precision of our finite size estimates.
The statistical precision we had on $E(L_N)$, up to $N\le 10^4$, 
was better than $0.002 \%$, and further improvement 
would have been very time consuming.

\begin{figure*}
\begin{center}
\resizebox{0.85\textwidth}{!}{\includegraphics{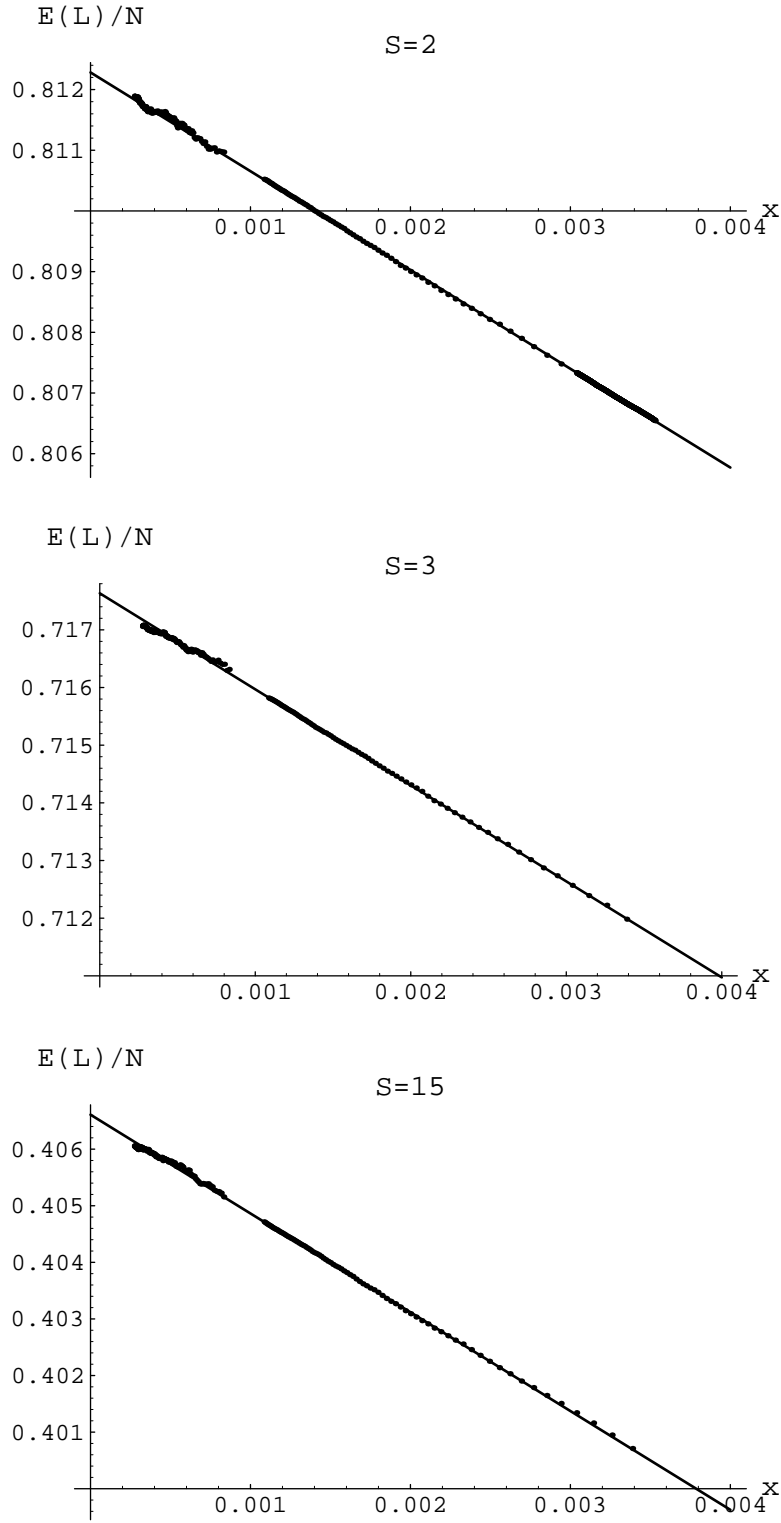}}
\end{center}
\caption{Extrapolation of the $N\le 10^4$ estimates for $E(L_N)/N$ 
to the large $N$ limit for $S=2,3$ and $15$. The solid 
curves represent best fits to a linear function of 
$x=1/(\ln{N}\sqrt{N})$. Estimates of $E(L_N)/N$ for $2.10^4\le N\le 10^5$, 
not taken into account in the extrapolation, are also included.} 
\label{fig_EL}
\end{figure*}

By using a best fit of our $N\le 1500$ estimates based on  
(\ref{fsize_EL}) with $\epsilon_N$ of the form
$\epsilon_N=K/(N\ln^{\alpha}{N})$ ($K$ and $\alpha$ being constants) 
we get a surprisingly good extrapolation up to values of $N$ of order 
$10^5$, which one would {\it not} be able to obtain 
by using another form than (\ref{fsize_EL}).

However the form chosen for $\epsilon_N$ remains somewhat 
arbitrary. Since the estimation of $\gamma_S$ and $A_S$ 
should be more precise when extrapolating from larger values of 
$N$, we performed a second series of extrapolations, using the 
finite size estimates obtained for $1500\le N\le 10^4$.
For these values of $N$, the term $\epsilon_N$ is much less 
significant, and a linear extrapolation of $E(L_N)$ 
as a function of $x=1/(\ln{N}\sqrt{N})$ is already very precise.
Figure (\ref{fig_EL}) reproduces our results in the cases 
$S=2$, $S=3$ and $S=15$. 
The solid curves in these figures are best fits of our 
$1500\le N\le 10^4$ estimates to a linear function of 
$1/(\sqrt{N}\ln{N})$.
In this way we obtained the estimates of $\gamma_S$ and $A_S$ 
which are given in table (\ref{tab_gammas}) a) for $2\le S\le 15$. 

\begin{table*}
\caption{Results of an extrapolation of our finite size 
estimates ($1500\le N\le 10^4$) based on (\ref{fsize_EL}) 
with different choices of $\epsilon_N$.
a) $\epsilon_N=0$;
b) $\epsilon_N\sim B_S/N$; 
c) $\epsilon_N\sim C_S/(N\ln{N})$.
The numbers in parentheses represent statistical 
errors obtained by $\chi^2$ analysis, in units of the last written digit.}
\label{tab_gammas}
\begin{center}
\begin{tabular}{|c|ccc|c|ccc|}
\multicolumn{8}{c}{a)}\\
\hline\noalign{\smallskip}
$S$&$\gamma_S$&$A_S$&- & $S$&$\gamma_S$&$A_S$&- \\
\hline\noalign{\smallskip}
2 & 0.812282(2) & -1.6276(5)& - & 9  & 0.493582(3) & -1.734(2) & -\\
3 & 0.717634(3) & -1.665(2) & - & 10 & 0.474702(2) & -1.742(1) & -\\ 
4 & 0.654304(11)& -1.677(7) & - & 11 & 0.458028(2) & -1.724(1) & -\\ 
5 & 0.607452(4) & -1.710(3) & - & 12 & 0.443168(3) & -1.721(2) & -\\ 
6 & 0.570625(3) & -1.729(2) & - & 13 & 0.429784(3) & -1.694(2) & -\\   
7 & 0.540509(2) & -1.729(1) & - & 14 & 0.417665(3) & -1.728(2) & -\\ 
8 & 0.515228(3) & -1.730(2) & - & 15 & 0.406609(4) & -1.745(3) & -\\
\hline\noalign{\smallskip}
\multicolumn{8}{c}{b)}\\
\hline\noalign{\smallskip} 
$S$&$\gamma_S$&$A_S$&$B_S$ & $S$&$\gamma_S$&$A_S$&$B_S$\\
\hline\noalign{\smallskip}
2 & 0.812386(4) & -1.765(5)& 0.59(2) & 9  & 0.493595(13) & -1.75(2) & 0.10(9)\\
3 & 0.717637(11)& -1.67(2) & 0.03(8) & 10 & 0.474696(9)  & -1.73(2) & -0.05(7)\\
4 & 0.654487(7) & -1.892(8)& 0.77(3) & 11 & 0.458017(9)  & -1.71(2) & -0.09(7)\\ 
5 & 0.607495(20)& -1.78(3) & 0.33(12)& 12 & 0.443176(12) & -1.73(2) & 0.06(9)\\
6 & 0.570658(12)& -1.78(2) & 0.25(8) & 13 & 0.429718(10) & -1.59(2) & -0.51(7)\\ 
7 & 0.540500(9) & -1.72(2) & -0.06(7)& 14 & 0.417627(13) & -1.67(2) & -0.3(1)\\ 
8 & 0.515173(10)& -1.64(2) & -0.42(8)& 15 & 0.406654(16) & -1.82(2) & 0.34(12)\\ 
\hline\noalign{\smallskip}
\multicolumn{8}{c}{c)}\\
\hline\noalign{\smallskip} 
$S$&$\gamma_S$&$A_S$&$C_S$ & $S$&$\gamma_S$&$A_S$&$C_S$\\
\hline
2 & 0.812370(3) & -1.726(3)& -2.95(10)& 9  & 0.493595(11)& -1.75(2) & -0.6(5)\\
3 & 0.717637(10)& -1.67(2) & -0.1(4)  & 10 & 0.474697(8) & -1.74(1) & 0.3(4)\\
4 & 0.654442(4) & -1.812(4)& -3.00(7) & 11 & 0.458019(8) & -1.71(1) & 0.4(4)\\
5 & 0.607490(14)& -1.76(2) & -1.8(7)  & 12 & 0.443175(10)& -1.73(2) & -0.3(5)\\
6 & 0.570653(10)& -1.77(2) & -1.3(5)  & 13 & 0.429728(8) & -1.62(1) & 2.7(4)\\
7 & 0.540502(8) & -1.72(1) & 0.4(4)   & 14 & 0.417635(11)& -1.69(2) & 1.5(5)\\
8 & 0.515182(9) & -1.67(1) & 2.2(4)   & 15 & 0.406649(14)& -1.80(2) & -1.9(7)\\
\hline\noalign{\smallskip}
\end{tabular}
\end{center}
\end{table*}

To obtain error bars on these estimates one should use a 
$\chi^2$ analysis \cite{BevingtonRobinson94_Book}. 
However this method underestimates the true error here.
Indeed, $\chi^2$ analysis leads to errors for the fitting parameters 
which decrease as $1/\sqrt{n}$ for $n>>1$, $n$ being the number of 
degrees of freedom, that is the number of {\it independent} datas
in the fit. Since we computed for each instance a whole profile, 
the averaged points in figure (\ref{fig_EL}) are not independent: 
There are correlations in the sequence $(L_i), 1\le i\le N$, which 
results in a smoothing of the averaged profile, or equivalently in 
a reduction of the ``effective'' number of independent datas in the fit. 
Moreover these correlations are long-ranged, which makes it uneasy to 
measure an effective number of degrees of freedom.
We thus relied on a semi-empirical method, 
by measuring the range over which the fitting parameters 
varied for different choices of $\epsilon_N$. 
Typically we obtained in this way an ``error'' less than 
$0.01\%$ on $\gamma_S$ and $5\%$ on $A_S$.
In fact $\epsilon_N$ happens to be only slightly larger than 
the precision of our finite size estimates for $1500\le N\le 10^4$.
We thus expect the above procedure to provide a faithful 
(slightly overestimated) measure of the true error on our estimates. 
Rather than quoting semi-empirical error bars, 
table (\ref{tab_gammas}) gives the results obtained by making
respectively the choices a) $\epsilon_N=0$, 
b) $\epsilon_N\sim B_S/N$ and c) $\epsilon_N\sim B_S/(N\ln{N})$.
Note that the cases b) and c) agree to a better accuracy together 
than with case a). To determine the precise form of the 
``second order'' corrections however clearly more precise computations 
would be needed.

\subsection{The variance of $L_N$ and the universality class of the
LCS problem.}

\begin{figure} 
\begin{center}
\resizebox{0.50\textwidth}{!}{\includegraphics{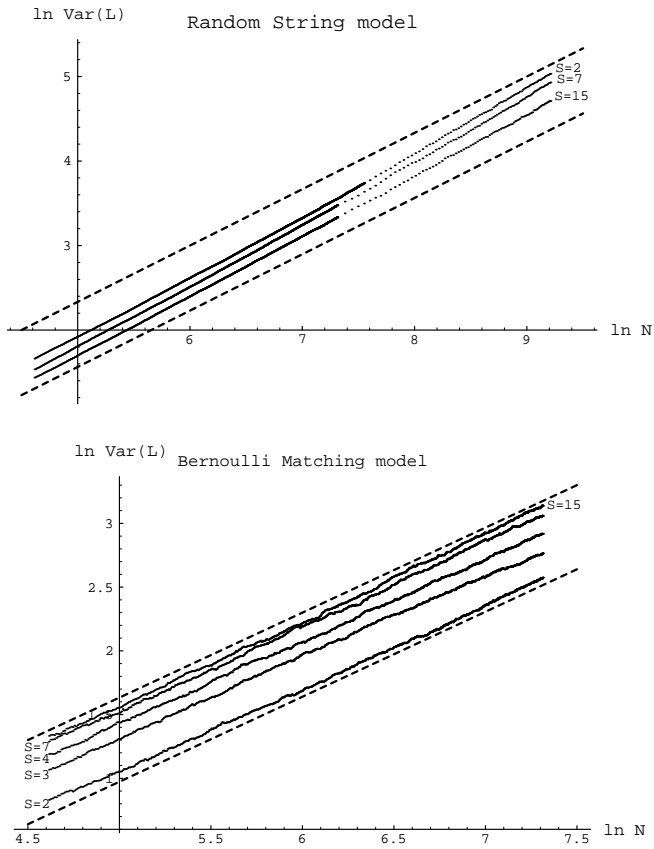}}
\end{center}
\caption{Scaling of the variance of the LCS length. 
Random String model: Averages over $10^5$ instances for 
$1\le N\le 1500$ and over $10^4$ instances for $1500\le 10^4$. 
Bernoulli Matching model: Averages over $10^4$ instances for
$1\le N \le 1500$. Dashed lines of slope $2/3$ give the 
expected scaling for the Bernoulli Matching model.}
\label{fig_varL}
\end{figure}

\begin{figure}
\begin{center}
\resizebox{0.40\textwidth}{!}{\includegraphics{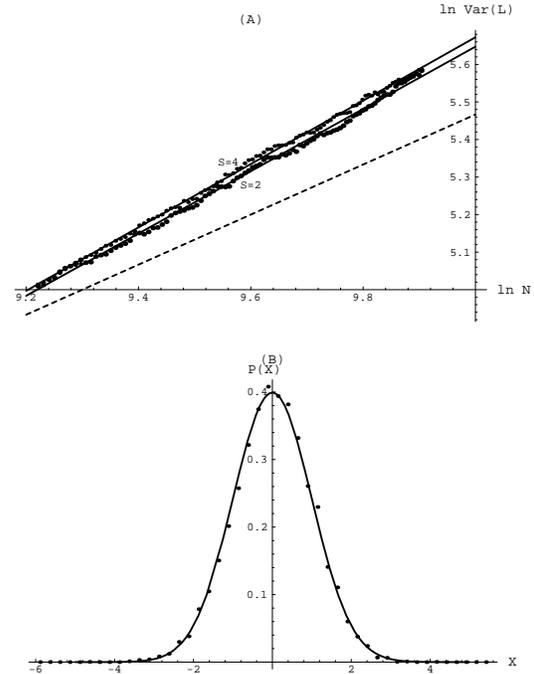}}
\end{center}
\caption{(A) Scaling of of $Var(L_N)$ for $10^4\le N\le 2.10^4$ 
in cases $S=2$ and $S=4$ (averages over $5000$ random strings).
The solid lines are best linear fits (slope $0.830$ for 
$S=2$ and $0.844$ for $S=4$). The dashed line has reference
slope $2/3$.
(B) Histogram of the values of $X_N$ for $S=2$ and $N=500$
(averages over $10^4$ random strings). 
The solid curve corresponds to the normal distribution with 
mean $0$ and unit variance.}
\label{fig_Lhist}
\end{figure}

It has been observed long ago by Chvatal and Sankoff 
\cite{ChvatalSankoff75_JAP} that the variance of the LCS length 
is numerically very small. These authors even conjectured that 
$Var(L_N)$ might be $o(N^{2/3})$. 
It has been suggested by Talagrand (in the context of longest increasing 
subsequences \cite{Talagrand96_AP}), that the smallness of $Var(L_N)$ 
may be related to the fact that the number of LCSs of two random 
sequences is very large. The only known general bound is however
$Var(L_N)=O(N)$, an immediate consequence of the concentration
inequality (\ref{conc_ineq}).
Anyway the work of Hwa and L\H{a}ssig \cite{HwaLassig96_PRL} provides 
a theoretical answer to the conjecture of Chvatal and Sankoff: 
The LCS problem falls into the universality class 
of a model of directed polymer in a 2D random potential. 
The variance of $L_N^B$ in the Bernoulli Matching model should grow 
as $N^{2/3}$. For the Random String model we must be cautious with 
this prediction, but it should provide at least a first 
approximation. The scaling behaviour of the variance of the LCS 
length is shown in figure (\ref{fig_varL}), both for the Random 
String model and for the Bernoulli Matching model, and for different 
values of $S$. We see as expected a very good agreement with the scaling 
law $Var(L_N^B)\approx N^{2/3}$ for the Bernoulli Matching model. 
The results for the Random String model are more interesting:
The scaling of $Var(L_N)$ is slightly, but clearly different from 
$N^{2/3}$. The correlations among the matches should be expected 
to be more and more relevant as $N$ grows, since $2N$ independant 
variables are involved in $L_N$ against $N^2$ in $L_N^B$.
In fact our results suggest that something like a crossover occurs 
from a small $N$ scaling regime where $Var(L_N)\approx N^{2/3}$ to 
an asymptotic scaling regime where $Var(L_N)\approx N^{2\omega'}$,  
$\omega'>1/3$. Note that this asymptotic regime seems not completely 
reached on figure (\ref{fig_varL}) which includes estimates for $N$ up 
to $10^4$. Hence the ``small $N$'' regime is rather extended. 
As is apparent on the figure, it becomes more and more extended as $S$ 
increases, and the asymptotic regime is more and more difficult to
reach. For this reason it is difficult to tell if the exponent $\omega'$ 
depends on $S$ or not. It is also difficult to tell if the numerical 
dependancies of $Var(L_N^B)$ and $Var(L_N)$ respectively on $S$ remain 
reversed in the asymptotic regime. As is seen on figure 
(\ref{fig_Lhist}A) however, our datas for $S=2$ and $S=4$ are almost 
indistinguishable in the range $10^4\le N\le 2.10^4$. We are thus 
tempted to conjecture that $\omega'$ is independent of $S$, and 
truly characterizes the universality class of the Random String model. 
Assuming that the asymptotic scaling regime 
is almost reached on figure (\ref{fig_Lhist}A) leads to the estimate 
$\omega'=0.418\pm 0.005$.

The {\it distribution} of $L_N$ is also of interest. 
We found that the random variable $X_N=(L_N-E(L_N))/\sqrt{Var(L_N)}$ is 
very nearly normally distributed even at rather small values of $N$.
These findings indicate that a central limit theorem 
should apply to the LCS length of two random strings, despite 
the nonlinear growth of $Var(L_N)$.
Figure (\ref{fig_Lhist}) shows the results of a computation
in the case of binary strings.

\subsection{Computations for the Bernoulli Matching model.}

We have performed similar Monte Carlo simulations for the 
Benoulli Matching model. 
For computational reasons (generating two pseudo-random strings 
of size $N$ is faster than a whole $N\times N$ matrix), 
we restricted extensive computations to sizes 
$N\le 1500$. We nevertheless performed a limited set of computations 
at sizes up to $N=10^5$, in order to check the validity of 
(\ref{fsize_EL}) in that case. We found that this finite 
size scaling law applies to the mean value $E(L^B_N)/N$ as well.
Using the same method as above we obtained the estimates of 
$\gamma^B_S$ which are quoted in table (\ref{tab_Bgammas}) 
for $2\le S\le 15$. These are not as precise as 
the corresponding estimates for the Random String model, since
the extrapolation was restricted to smaller values of $N$.
However we estimate the precision on $\gamma^B_S$ to be 
better than $0.1\%$. More interestingly, we found that 
the values of $\gamma^{B}_S$ are very well reproduced 
by the simple expression 
\begin{equation} \label{gammaB_conjecture}
\gamma^{B}_S=2/(1+\sqrt{S}), 
\end{equation}
a formula which had already been conjectured by 
Steele \cite{Steele82_SIAMJAM,Steele97_Book}. 
In fact Steele made his conjecture for the original 
LCS problem, at a time where precise numerical estimates 
of $\gamma_S$ where not available, but it happens to 
be valid for the Bernoulli Matching model. 

\begin{table}
\caption{Estimates of $\gamma^B_S$ for $2\le S\le 15$.
The extrapolation of $E(L^B_N)/N$, $N\le 1500$, was based 
on (\ref{fsize_EL}) with $\epsilon_N\sim C_S/(N\ln{N})$ 
(values obtained for $A_S$ and $C_S$ are not reproduced). 
Precision on $\gamma_S$, estimated as the range of variation 
of our estimates for several choices 
$\epsilon_N\sim K_S(N\ln^{\alpha}{N})^{-1}$, is about $0.05\%$. 
The conjectured values $2/(1+\sqrt{S})$ for $\gamma^B_S$ 
are also quoted.}
\label{tab_Bgammas} 
\begin{tabular}{|c|cc|c|cc|}
\hline
$S$&$\gamma^B_S$&$2/(1+\sqrt{S})$&$S$&$\gamma^B_S$&$2/(1+\sqrt{S})$\\
\hline
2 & 0.82860 & 0.828427 & 9  & 0.50047 & 0.5\\  
3 & 0.73236 & 0.732051 & 10 & 0.48082 & 0.480506\\ 
4 & 0.66698 & 0.666667 & 11 & 0.46383 & 0.463325\\ 
5 & 0.61823 & 0.618034 & 12 & 0.44850 & 0.448018\\ 
6 & 0.58030 & 0.579796 & 13 & 0.43484 & 0.434259\\ 
7 & 0.54892 & 0.548584 & 14 & 0.42223 & 0.421793\\ 
8 & 0.52291 & 0.522408 & 15 & 0.41077 & 0.410426\\
\hline
\end{tabular}
\end{table}

\begin{figure} 
\begin{center}
\resizebox{0.40\textwidth}{!}{\includegraphics{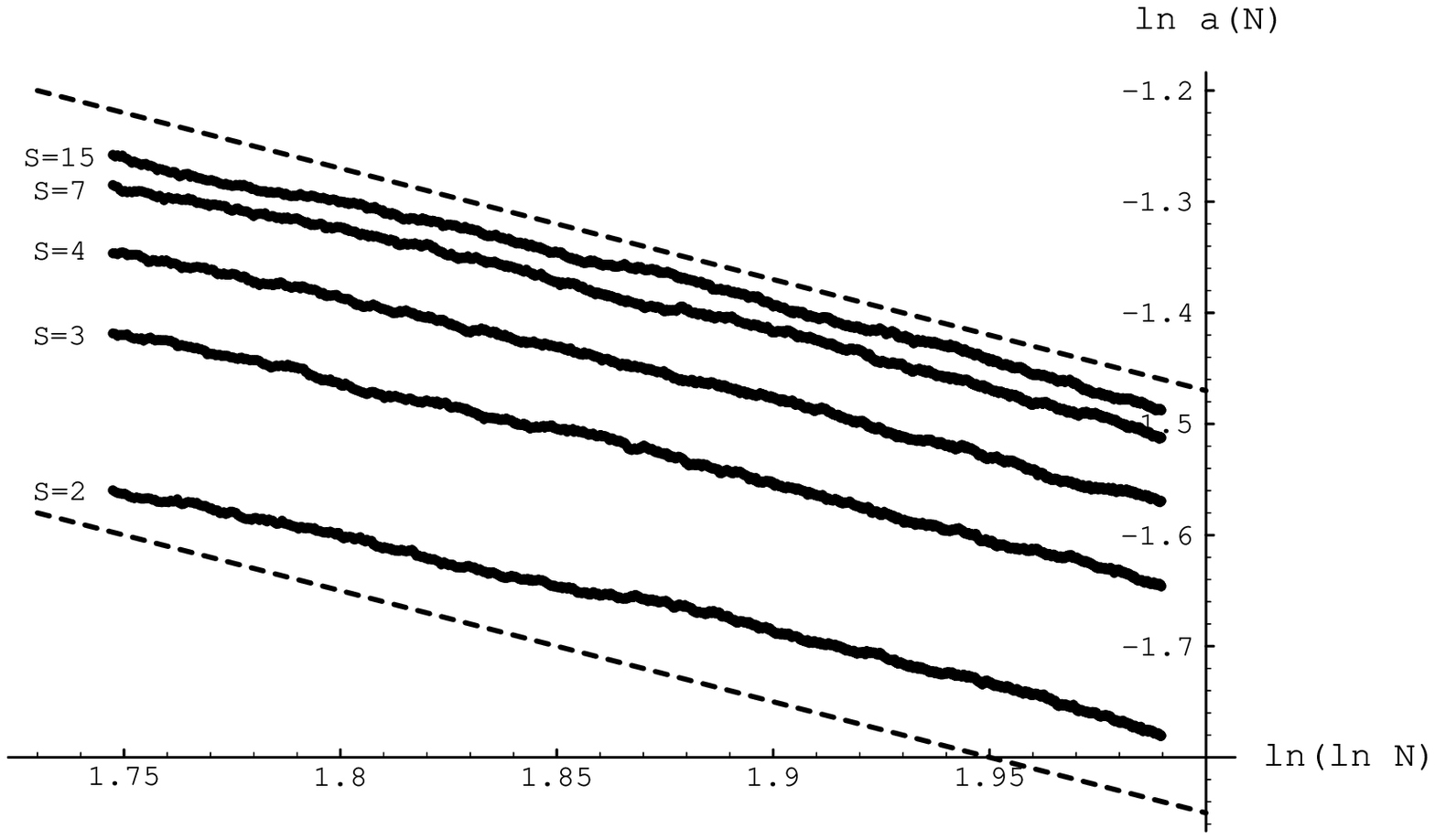}}
\end{center}
\caption{Scaling of the finite size corrections to linear growth 
for the Bernoulli Matching model. The figure represents a 
plot of $\ln{a(N)}$ defined in the text in function of $\ln{\ln N}$ 
($300\le N\le 1500$) for different values of $S$. Dashed lines with slope 
$-1$ visualize the scaling expected from (\ref{fsize_EL}).}
\label{fig_loga}
\end{figure}

\bigskip
A short discussion may be instructive. Let $A_k$ 
be the event that there exists a sequence of matches of 
length $k$. Then the length of a longest sequence of matches is
\begin{equation}
L= \sum_{k=1}^N 1_{A_k},
\end{equation}
where $1_A$ is the indicator of set $A$ in the sample 
space $\Omega$ of the model (be it the random string 
model or the Bernoulli Matching model).
Hence the mean value $E(L)$ essentially depends on the behaviour of  
the probabilities $P(A_k)$: Using the martingale difference method 
(see e.g. \cite{Steele97_Book}), one finds that 
\begin{equation} \label{conc_ineq}
P(|L-E(L)|\ge k)\le 2e^{-{k^2\over 8N}}
\end{equation}
hence $P(A_k)\ge 1-2\exp(-(EL-k)^2/8N)$ for $k\le E(L)$, and 
$P(A_k)\le 2\exp(-(k-EL)^2/8N)$ for $k\ge E(L)$. 
 
The location of this transition is very difficult 
to compute, but it is clearly related to the behaviour of 
the random variable ${\cal N}_k(\omega)$ defined as the number 
of sequences of matches of length $k$ for a given instance 
$\omega$. 
Clearly
\begin{equation} \label{upper_PAk}
P(A_k)\le E({\cal N}_k) = S^{-k} {N\choose k}^2.
\end{equation} 
Setting $k=xN$ for $0<x<1$ and using Stirling formula, it is found 
\cite{ChvatalSankoff75_JAP} that $E({\cal N}_k)$ has a transition from 
exponentially growing to exponentially decreasing behaviour at a
value $x={\hat x}_S$ given by the solution to 
$x(1-x)^{(1-x)/x}=S^{-1/2}$.
Hence ${\hat x}_S$ is an upper bound for 
$\gamma_S$ and $\gamma^{B}_S$. It is not very accurate: 
One has ${\hat x}_2\approx 0.9$, and as $S\to \infty$,
${\hat x}_S\sim e/\sqrt{S}$ which is not what one would 
expect from (\ref{gammaB_conjecture}).
The reason of this failure is that ${\cal N}_k$ is
not a self-averaging quantity, so that its mean value does 
not reproduce well its typical behaviour. Consider then the 
``entropy'' $\ln ({\cal N}_k+1)$. This is a self-averaging quantity
from which $\gamma_c$($=\gamma_S$ or $\gamma^{B}_S$) can be computed
as the smallest number $0<\gamma<1$ such that $x>\gamma$ implies
\begin{equation}
\lim_{N\to \infty} {E\ln ({\cal N}_{xN}+1) \over N}=0.
\end{equation}
Clearly the function 
$f(x)=\lim_{N\to \infty} N^{-1}E\ln({\cal N}_{xN} +1)$ 
is singular at $x=\gamma_c$. From the results of
section (\ref{config_space_properties}), we even expect 
$f(x)$ to be discontinuous at $x=\gamma_c$. 
Unfortunately, computing $E\ln({\cal N}_k+1)$ is still a 
difficult problem. 

Steele suggested another approach to the problem 
\cite{Steele98_private}, which consists of looking 
at the {\it maximum} of ${\cal N}_k(\omega)$. 
The location $k_{max}$ of this maximum is a self-averaging 
quantity which may be expected to be comparable 
in a simple way with the LCS length: A plausible guess 
is that with probability one, $k_{max}/L\to 1/2$ as $N\to \infty$.
Assuming this we must maximize $f(x)$ defined 
above, and the situation is not much better than before.
But now the approximation of replacing ${\cal N}_k$ by 
its mean value does work much better:
$E({\cal N}_k)$ has a sharp maximum at 
$k\sim x_S N$, where $x_s=1/(1+\sqrt{S})$. 
Hence quite surprisingly, $2x_S$ is a really good 
estimate for $\gamma_S$, and it happens to give the correct 
value of $\gamma^{B}_S$. We have no explanation for this 
observation, but we remark that a  
similar computation can be done for the related Longest Increasing 
Subsequence (LIS) Problem. Given a sequence of distinct numbers 
$x_1,...,x_N$ this problem asks for a sequence 
$1\le i_1<...<i_k\le N$ such that $x_{i_1}<...<x_{i_k}$ and k 
is maximal. When the $x_i$'s are i.i.d. random variables uniformly 
distributed in $[0,1]$, it is known that the expected length of a 
LIS is asymptotic to $\gamma_{IS} \sqrt{N}$ where 
$\gamma_{IS}=2$ \cite{VersikKerov77_SMD}. 
Now let ${\cal N}^{(IS)}_k$ be the number of increasing 
subsequences of length $k$ of $x_1,...,x_N$, so that 
\begin{equation}
E({\cal N}^{(IS)}_k)={N\choose k} {1\over k!}.
\end{equation}
Using Stirling formula, one finds that
$E({\cal N}^{(IS)}_k)$ has a transition from a rapidly growing 
to a rapidly decreasing behaviour at $k\sim e\sqrt{N}$, and
presents a sharp maximum at $k \sim x_{IS} \sqrt{N}$ where $x_{IS}=1$. 
Hence $\gamma_{IS}=2x_{IS}$ and the above 
approximation is actually exact in this case.

As a byproduct, expression (\ref{gammaB_conjecture}) provides
a consistent mean to check the validity of the finite size scaling 
(\ref{fsize_EL}) for the Bernoulli Matching model. Indeed we can 
measure directly the scaling in $N$ of the quantity
\begin{equation}
a_S(N)=\sqrt{N}(\gamma_S^B N-E(L_N^B)).
\end{equation}
As is shown in figure (\ref{fig_loga}), $\ln{a_S(N)}$ has a near linear 
dependance on $\ln(\ln N)$ with a slope consistent with $-1$,
as is expected by assuming the validity of (\ref{fsize_EL}).

\section{The case $N\ne M$ and a cavity solution.}
\label{N_ne_M_case}

There is still another way to study the asymptotic 
behaviour of $E(L_N)$, which consists of working 
directly with the recurrence relation (\ref{Lij_rec}).
This point of view has the advantage that it enables one 
to study the case $M\ne N$ in a natural way, leading
to a generalization of (\ref{gammaB_conjecture}) to the case
where $N,M\to \infty$, the ratio $r=M/N$ being fixed.

In order to find the asymptotics of (\ref{Lij_rec}) it is 
convenient not to work with $L_{ij}$ directly, but rather 
(as in \cite{MasekPaterson80_JCSS}) with the differences 
$\nu_{ij}$ and $\mu_{ij}$ defined by
\begin{equation}
\nu_{ij}=L_{ij}-L_{i-1,j}, \mu_{ij}=L_{ij}-L_{i,j-1},
\indent 1\le i,j\le N.
\end{equation}
The recurrence relations for $\nu_{ij}$ and $\mu_{ij}$ are readily
seen to be
\begin{eqnarray}\label{nu_mu_rec}
\nu_{ij}=\max\big(0,\epsilon_{ij}-\mu_{i-1,j},
\nu_{i,j-1}-\mu_{i-1,j}\big) \nonumber\\
\mu_{ij}=\max\big(0,\epsilon_{ij}-\nu_{i,j-1},
\mu_{i,j-1}-\nu_{i,j-1}\big)
\end{eqnarray}
with boundary conditions $\nu_{i,0}=\nu_{0,i}=\mu_{i,0}=\mu_{0,i}=0$.
In the Random String model we have $\epsilon_{ij}=\delta_{X_i,Y_j}$,
whereas in the Bernoulli Matching model the $\epsilon_{ij}$'s are 
i.i.d. Bernoulli variables with 
$P(\epsilon_{ij}=1)=1-P(\epsilon_{ij}=0)=1/S$.
We consider relations (\ref{nu_mu_rec}) as a kind of 
exact cavity equations \cite{MezardParisiVirasoro87_Book} 
for the LCS problem.
The LCS length $L_N$ can be retrieved by summing the $\nu_{ij}$'s and 
$\mu_{ij}$'s along the first bissector. To be precise
\begin{equation}
L_N=\sum_{i=1}^N \big(\nu_{ii}+\mu_{i-1,i}\big)
=\sum_{i=1}^N \big(\mu_{ii}+\nu_{i,i-1}\big).
\end{equation}
When $N\ne M$, $M/N=r$, we view $L_{NM}$ as a sum along a path 
``as straight as possible'' in the direction defined by $r$,
e.g. a path zigzaging along the straight line 
joining the points $(0,0)$ and $(N,M)$ in such a 
way as to keep as close as possible from this line.

A simple, but important observation is that the variables $\nu_{ij}$
and $\mu_{ij}$ can take only the values $0$ and $1$. Hence let us
introduce the probabilities
\begin{equation}\label{pij_lim}
p_{ij}=P(\nu_{ij}=1),\indent p'_{ij}=P(\mu_{ij}=1).
\end{equation}
As $i$ and $j\rightarrow \infty$, it is natural to expect that 
$p_{ij}$ and $p'_{ij}$ have limits depending only on the ratio 
$r=j/i$. We denote these limits respectively by $p(r)$ and $p'(r)$
(or $p$ and $p'$ for short). 

For a given $r>0$, $L_{N,[rN]}$ is a sum of $N$ 
terms $\nu_{ij}$ and $rN$ terms $\mu_{ij}$, along a path 
``close'' to the straight line from $(0,0)$ to $(N,[rN])$.
Hence the limit $\gamma_S(r)=\lim_{N\to \infty} E(L_{N,[rN]})/N$ 
must be given by
\begin{equation}\label{gamma_p_p'}
\gamma_S(r)=p(r) + rp'(r).
\end{equation}
Now a relation between $p(r)$ and $p'(r)$ can be readily obtained
from (\ref{nu_mu_rec}) if we assume that for large $i$ and $j$, 
the occupation numbers $\nu_{i,j}$ and $\mu_{i,j}$ are
nearly independent variables. It turns out that this decorrelation 
property holds true in the Bernoulli Matching model. It can be justified
on the basis of a transfert matrix method for this percolation problem
which will be presented elsewhere \cite{Boutet98_PRL}.
In the limit $i,j\to \infty$ we are thus led to the 
following self-consistent equations:
\begin{eqnarray}\label{cavity_lcs}
p_{ij}=1-p'_{i-1,j}-(1-1/S)(1-p_{i,j-1})(1-p'_{i-1,j}),\nonumber \\ 
p'_{ij}=1-p_{i,j-1}-(1-1/S)(1-p_{i,j-1})(1-p'_{i-1,j}).
\end{eqnarray}
If we now let $p(r)=\lim_{i\to\infty} p_{i,[ri]}$ and 
$p'(r)=\lim_{i\to\infty} p'_{i,[ri]}$ and note that 
$p_{i,[ri]-1}=p_{i,[ri]}-1/i(d/dr)p(r)$ and 
$p'_{i-1,[ri]}=p'_{i,[ri]}+r/i(d/dr)p'(r)$ up to negligible
terms in the limit $i\to \infty$, then taking the sum and the 
difference in (\ref{cavity_lcs}) leads to
\begin{equation} \label{p_p'_equation}
1=p+p'+(S-1)pp'
\end{equation}
and
\begin{equation} \label{dp_dp'_equation}
{d\over dr}p(r) + r{d\over dr}p'(r) = 0.
\end{equation}
These last two equations determine the functions $p(r)$ and
$p'(r)$ completely. A simple computation gives now
\begin{equation}\label{p_p'_expression}
p(r)={\sqrt{rS}-1\over S-1}, \indent p'(r)={\sqrt{S/r}-1\over S-1}.
\end{equation}
Note that the relation $p'(r)=p(1/r)$ is obvious from symmetry 
considerations. It must be also remarked that (\ref{p_p'_expression}) 
is only satisfied for $1/S\le r \le S$ (although (\ref{p_p'_equation}) 
and (\ref{dp_dp'_equation}) are valid for all $r$ except $r=S$ and $r=1/S$):
The LCS problem has a percolation transition when one of the two 
strings is $S$ times larger than the other. 
Suppose for instance $r=M/N=S$. 
Consider the sequence of matches $(1,j_1),(2,j_2),...$, 
where $j_1$ is the smallest integer $j\ge 1$ such that $(1,j)$ 
is a match, $j_2$ is the smallest integer $j>j_1$ such that $(2,j)$ 
is a match, and so on. The differences $j_{k+1}-j_k$ are 
independent random variables with mean value $S$. By the law of 
large numbers, $j_k$ is asymptotic to $kS$ as $k\to \infty$.
It follows that the length of this sequence of matches, restricted
to the integer points $(ij)$ such that $j\le M$, is asymptotic
to $M/S=N$ as $N\to \infty$. Hence for $r\ge S$ we have
$\gamma_S(r)=\gamma^B_S(r)=1$ (and also $\gamma_S(1/r)=\gamma^B_S(1/r)=1/r$ 
by symmetry). This means that when $i$ is large and $j\ge Si$, $L_{ij}$ is
nearly equal to $i$, hence for each $i'\le i$ and $j'\ge j$ 
we have $\nu_{i'j'}=1$ and $\mu_{i'j'}=0$ with high probability.
In other words, $r\ge S$ implies $p(r)=1$ and $p'(r)=0$, and 
by symmetry, $p(1/r)=0$ and $p'(1/r)=1$.
From (\ref{p_p'_expression}) and (\ref{gamma_p_p'}) we find the expression
of the function $\gamma^B_S(r)$ of the Bernoulli Matching model for 
$1/S\le r\le S$:
\begin{equation}\label{gammaBr_expression}
\gamma^B_S(r)={2\sqrt{rS} - r-1\over S-1}.
\end{equation}
Note that the transition of $\gamma^B_S(r)$ at $r=S$ and $r=1/S$ is 
``second order'', that is $d\gamma^B_S/dr=p'(r)$ is continuous and 
$d^2\gamma^B_S/dr^2(r)$ is discontinuous at $r=S$ and $r=1/S$. 

\begin{figure} 
\begin{center}
\resizebox{0.50\textwidth}{!}{\includegraphics{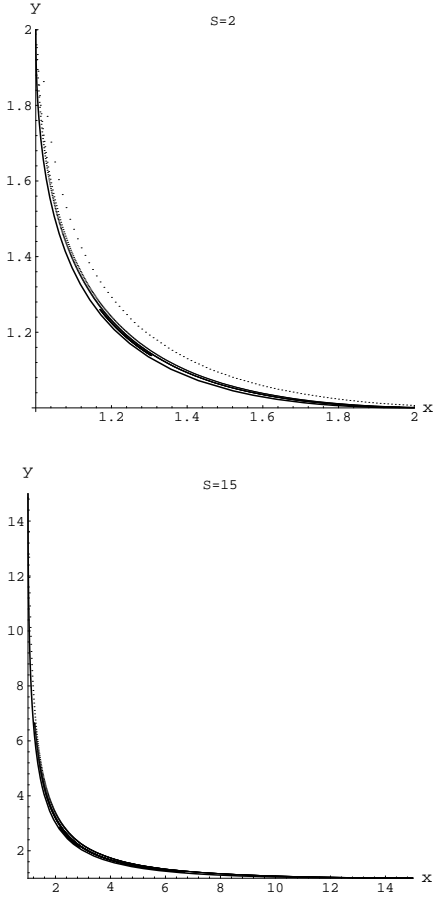}}
\end{center}
\caption{The boundary shape of the set $C_t/t$ for the Bernoulli
Matching model for $S=2$ ($t=100,500,1000,2300$) and 
$S=15$ ($t=100,300,700$). 
Each dotted line represents an average over $1000$ instances
of size $N$ ($(N,N)$ Bernoulli matrices) with $N=3000$ for $S=2$ 
and $N=2000$ for $S=15$.   
The solid curve is the asymptotic shape predicted from 
(\ref{gammaBr_expression}).}
\label{fig_bmphi}
\end{figure}

Figure (\ref{fig_bmphi}) shows the confrontation of equation 
(\ref{gammaBr_expression}) to a Monte Carlo computation of the 
Bernoulli Matching model for $S=2$ and $S=15$. 

\begin{figure} 
\begin{center}
\resizebox{0.50\textwidth}{!}{\includegraphics{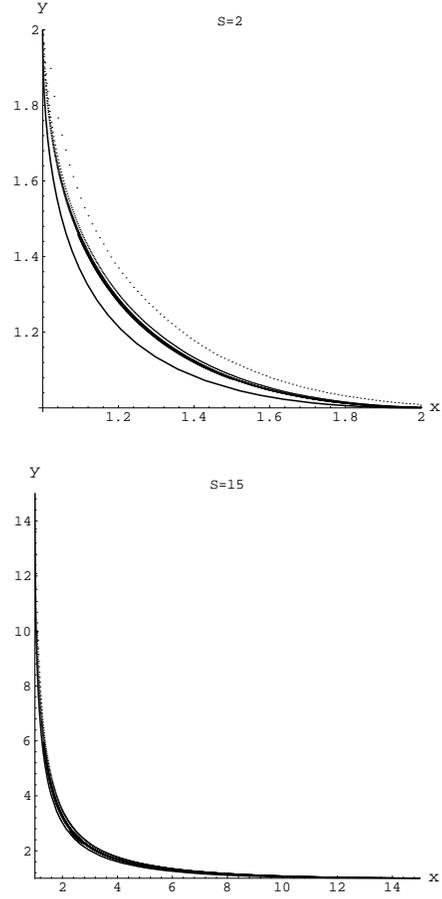}}
\end{center}
\caption{The boundary shape of the set $C_t/t$ for the LCS problem
(Random String model) for $S=2$ ($t=100,500,1000,1500,2000$)  
and $S=15$ ($t=100,300,500,1100$). Each dotted line represents 
an average over $1000$ random strings of size $N=3000$.
The solid curve, plotted for comparison, is the asymptotic 
shape predicted from (\ref{gammaBr_expression}) for the Bernoulli 
Matching model.} 
\label{fig_lcsphi}
\end{figure}

We have plotted, for several values of $t$, 
the ``curves'' delimiting the set $C_t/t$ in the two dimensional 
(x,y) plane, where $C_t=\{(ij): 1\le i,j\le N, L_{ij}\le t\}$. 
As $t\to \infty$, the boundary of $C_t/t$ approaches asymptotically 
the curve of parametric equation 
$r\to (1/\gamma^B_S(r),r/\gamma^B_S(r))$. This is the solid curve 
which we have plotted using (\ref{gammaBr_expression}). 
Figure (\ref{fig_lcsphi}) reproduces for comparison the 
results of analogous computations made for the Random String model.
Note that as $S$ increases, the differences between the results for
the Bernoulli Matching model and the random string model are less and 
less significant, and it is reasonable to 
expect that $\gamma_S(r)$ is asymptotic to $\gamma^B_S(r)$ 
as $S\to \infty$. Numerically the convergence is rather rapid:
the quantity $S(\gamma^B_S-\gamma_S)$ shows a maximum 
at $S\approx 11$ after which it happens to decrease. 
Such a phenomenon has already been observed 
and interpreted in other combinatorial optimisation problems 
\cite{BoutetMartin97_PRL}, and it would be of interest to have a 
theoretical understanding of the large $S$ behaviour of 
$\gamma^B_S-\gamma_S$. We leave this question open 
for future work.

\section{Configuration space properties of the LCS problem.}
\label{config_space_properties}

In this section we study generic properties of the set of 
solutions of the LCS problem, that is average properties of 
the set of all LCSs of two random strings.

A most direct computational access to these properties is provided by 
what we shall call the LCS graph of a given instance.
Given any strings $X$ and $Y$ of length $N$, this graph is defined as
follows. The vertices are the {\it LCS matches}, that is the set of points 
$(ij)$, $1\le i,j\le N$ such that $X_i=Y_j$ and $(ij)$ occurs in at least 
a LCS of $X$ and $Y$. Two LCS matches are incident in the LCS graph if 
they occur as successive matches (regardless the order) in the same LCS. 

It is a nice feature of the LCS problem that this structure may
be computed in a very efficient way. 
To a large part, this circumstance is due to the directed 
nature of the problem, which greatly simplifies the structure 
of the set of solutions.

\bigskip
\subsection{Construction of the LCS graph.} 
\label{LCSgraph_construction}

\bigskip
Since the construction we have used is rather simple we shall 
not give a precise algorithm, but rather indicate the main steps, 
together with the main observations which enable an efficient 
implementation.

Given integer points $(i_1j_1)$ and $(i_2j_2)$ we write 
$(i_1j_1)\prec (i_2j_2)$ if $i_1<i_2$ and $i_2<j_2$. 
Suppose the LCS matrix of $X$ and $Y$ is computed, and let 
$L$ be the length of a LCS of $X$ and $Y$. 
Following the terminology of \cite{ApostolicoGuerra87_Alg}, we call
an integer point $(ij)$ such that $X_i=Y_j$ a match of 
{\it rank} $k$ if $k$ is the length of a LCS of 
$X_1,...,X_i$ and $Y_1,...,Y_j$. 
It is then easy to construct, for each $1\le k\le L$, a list $M(k)$
of the matches of rank $k$ of $X$ and $Y$.
It is convenient to have the members $(ij)$ of $M(k)$ ordered 
lexicographically, in such a way that $i$ and $j$ vary in 
{\it opposite} directions, e.g. $i$ increasing while $j$ 
is decreasing.
Then setting $M(k)=\{(i_1j_1),...,(i_{m_k}j_{m_k})\}$,
one sees that $(i_1,...,i_{m_k})$ is an increasing sequence,
while $(j_1,...,j_{m_k})$ is a decreasing sequence.
The reason for this is that given any two members $(ij)$ 
and $(i'j')$ of $M(k)$ we have $i<i' \Rightarrow j\ge j'$, 
since otherwise $(ij)$ and $(i'j')$ would not be of the same 
rank. This property is important for an efficient 
construction of the LCS graph.
 
The lists $M(k)$ are the basic data in the construction of 
the LCS graph.
Remark that the members of $M(L)$ are obviously LCS matches,
hence these must be included as vertices of the LCS graph.
If $P$ is a match of rank $k<L$, then $P$ is a LCS match
if and only if there is a LCS match $Q$ of rank $k+1$ such 
that $P\prec Q$. Remark also that, by definition, a LCS match of rank 
$k$ may be connected only to LCS matches of rank $k-1$ or $k+1$ in 
the LCS graph. If $P$ is a LCS match of rank $k>1$, and $Q$ 
is a LCS match of rank $k-1$, then $P$ and $Q$ are connected 
if and only if $Q\prec P$. We will denote by $M_{LCS}(k)$ the list of
the LCS matches of rank $k$, ordered in the way which is inherited 
from the ordering of $M(k)$.

We construct the LCS graph in $L$ stages numbered 
$k=L,L-1,...,1$. Stage $L$ consists of inserting all matches of 
rank $L$ as vertices of the LCS graph. Once all the LCS matches 
of rank $>k$ have been inserted, stage $k$ consists of selecting 
the members of $M(k)$ which belong to $M_{LCS}(k)$, and then
to insert the required edges connecting $M_{LCS}(k)$ to 
$M_{LCS}(k+1)$.

Using remarks made previously and exploiting the way $M(k)$ and 
$M_{LCS}(k+1)$ have been ordered, it is easy to see that the 
selection of the members of $M_{LCS}(k)$ from those of $M(k)$ 
at stage $k$ may be performed in $O(m_k+l_{k+1})$ steps, 
$m_k$ and $l_{k+1}$ being the cardinality of $M(k)$ and 
$M_{LCS}(k+1)$ respectively. 
Hence the detection of the {\it whole} set of 
LCS matches takes at most $O(m)$ steps in this construction, 
$m=\sum_k m_k$ being the total number of matches of 
$X$ and $Y$.
The main part of the computation is devoted to the insertion 
of the edges in the LCS graph. The number of operations 
(comparisons and insertions) needed to determine the edges connecting 
$M_{LCS}(k)$ and $M_{LCS}(k+1)$, once these lists are known, 
is of order $O(l_k^2)$.
Since there is no obvious bound for $l_k$ better than $m_k$, 
and no obvious bound for $m_k$ better than $2N$, we obtain a bound
for the time required to compute the LCS graph which is $O(LN^2)$.

However when $X$ and $Y$ are random strings from a finite 
alphabet, the typical values of $l_k$ happen to be much 
smaller than $m_k$, and the typical time required by the above
construction is in fact much smaller than $O(LN^2)$.

\bigskip
\subsection{Computations of the LCS graph.} 
\label{LCSgraph_computations}

\bigskip
We performed a series of Monte Carlo computations 
in order to study some of the basic properties of the set of 
LCSs of two random strings. 
We concentrated our study on different quantities which 
can be easily computed once the LCS graph is constructed.

\begin{figure} 
\begin{center}
\resizebox{0.50\textwidth}{!}{\includegraphics{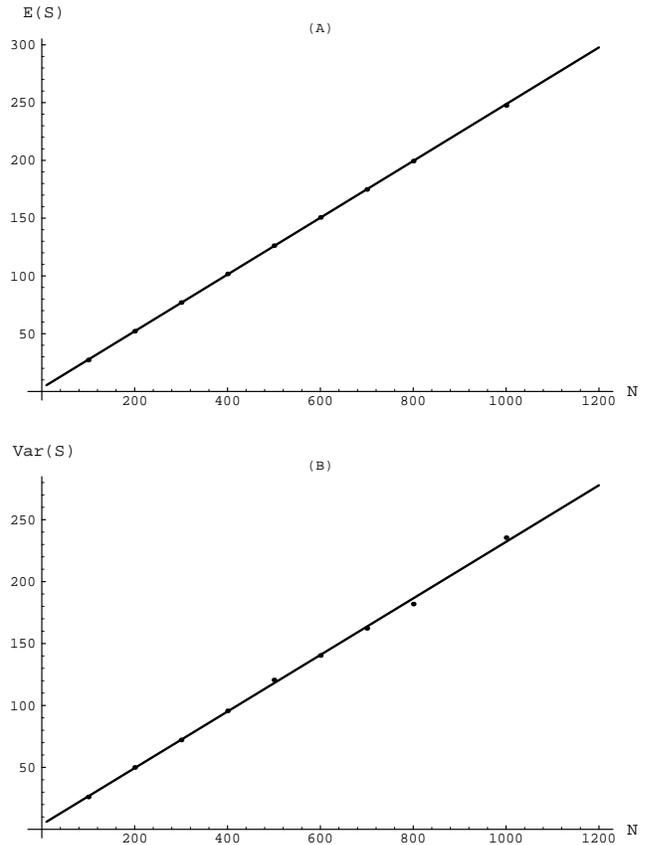}}
\end{center}
\caption{Mean value (A) and variance (B) of the ground state 
entropy ${\cal S}_N=\ln {\cal N}_{LCS}$ as a function of $N$, for $S=2$
(Random String model, averages over $10^4$ instances).}
\label{fig_Nlcs}
\end{figure}

Probably the most basic quantity which characterizes the
set of LCSs is its cardinality ${\cal N}_{LCS}$. Figure 
(\ref{fig_Nlcs}) reproduces the estimated
average and variance of the ground state entropy 
${\cal S}_N=\ln {\cal N}_{LCS}$ in case $S=2$, computed over $10^4$ 
random instances and for values of $N$ ranging from $100$ to $1000$. 
It is rather striking on this figure that $E({\cal S}_N)$ grows 
linearly with $N$. 
We expect the random variable $\ln {\cal N}_{LCS}$ to 
be self-averaging, and this is confirmed by the 
measured behaviour of its variance, whose growth is also
nearly linear. We observed this behaviour for all 
the values of $S$ we considered.

Hence we found that the number of LCSs of two typical
random strings is very large. ${\cal N}_{LCS}$ 
typically grows exponentially with $N$, with a well defined 
exponential factor $\alpha_S$, which we define, assuming 
the limit indeed exists, as
\begin{equation}
\alpha_S= \lim_{N\to \infty} {E({\cal S}_N)\over N}.
\end{equation}
Also we define (provided the limit exists)
\begin{equation}
\beta_S= \lim_{N\to \infty} {Var({\cal S}_N)\over N}.
\end{equation}
Using best linear fits we obtained rather precise estimates 
of $\alpha_S$ and $\beta_S$, which are quoted in table 
(\ref{tab_alpha_q}) for several values of $S$.

Another quantity reflecting the ``size'' of the set of LCSs of 
two random strings is the typical overlap of two LCSs.
Viewing a LCS of $X$ and $Y$ as a 
sequence of integer points we define the
overlap of two LCSs $\sigma_1=(Q_1,...,Q_L)$ and 
$\sigma_2=(P_1,...,P_L)$ as the quantity 
\begin{equation}
q=q(\sigma_1,\sigma_2)={1\over L} \sum_{k=1}^L \delta(Q_k,P_k).
\end{equation}
where $\delta(Q_k,P_k)=1$ if $Q_k=P_k$ and $0$ otherwise.
$q(\sigma_1,\sigma_2)$ is analogous to the order parameter 
used in the theory of spin glasses \cite{MezardParisiVirasoro87_Book}.
The quantity $L(1-q(\sigma_1,\sigma_2))$ should be regarded as 
a kind of Hamming distance in the space of LCSs of
$X$ and $Y$. The object of interest here is the empirical 
distribution of $q(\sigma_1,\sigma_2)$ for $\sigma_1$ and 
$\sigma_2$ ranging over the set of LCSs of $X$ and $Y$. We denote 
by $<q>$ and $<q^2>$ the first and second moment of the overlap
under this distribution. It is not difficult to see that
\begin{equation}
<q> = {1\over L} \sum_{k=1}^L \sum_{Q\in M_{LCS}(k)} P_1(Q)^2
\end{equation}
where 
\begin{equation}
P_1(Q)={{\cal N}_{LCS}(Q)\over {\cal N}_{LCS}},
\end{equation}
and ${\cal N}_{LCS}(Q)$ is the number of LCSs of $X$ and $Y$ of 
which the integer point $Q$ is a member.
Hence the average overlap $<q>$ can be easily computed for any given
instance of $X,Y$ once the LCS graph is constructed.
Also we have
\begin{equation}
<q^2>={1\over L^2} \sum_{k=1}^L \sum_{l=1}^L \sum_{Q\in M_{LCS}(k)}
\sum_{Q'\in M_{LCS}(l)} P_2(Q,Q')^2
\end{equation}
where
\begin{equation}
P_2(Q,Q')={{\cal N}_{LCS}(Q,Q')\over {\cal N}_{LCS}}
\end{equation}
and ${\cal N}_{LCS}(Q,Q')$ is the number of LCSs of $X$ and $Y$ 
of which points $Q$ and $Q'$ are members. It is still 
elementary to compute $<q^2>$, but more computationally
lengthy due to the above double summation.

\begin{figure} 
\begin{center}
\resizebox{0.50\textwidth}{!}{\includegraphics{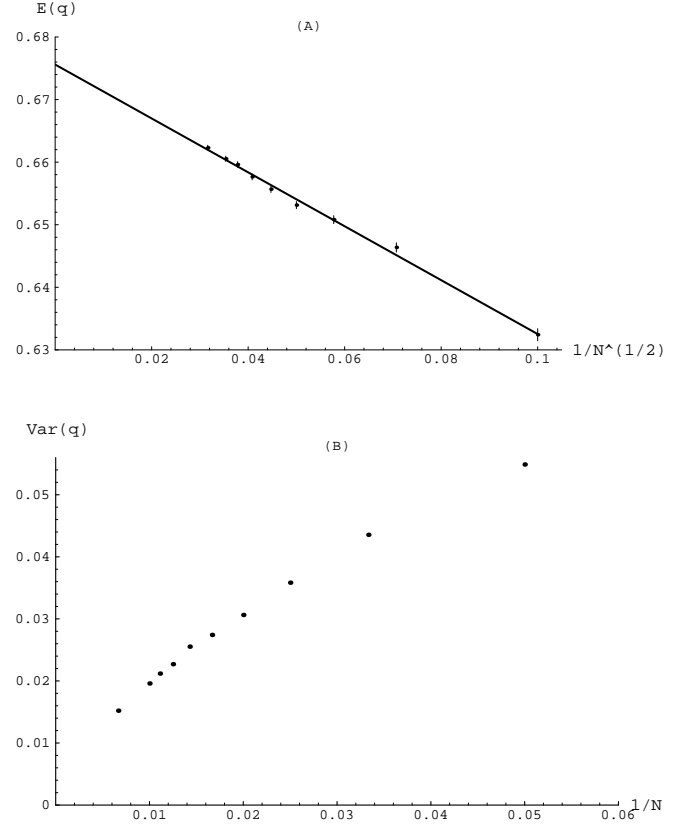}}
\end{center}
\caption{(A) The average overlap $E(q)$ of two random LCSs as a function
of $1/\sqrt{N}$ ($100\le N\le 1000$, averages over $10^4$ random strings). 
(B) The variance $Var(q)=E(q^2)-(Eq)^2$ of $q$ as a function of $1/N$ 
($10\le N\le 100$). Statistical error bars in (A) were obtained from estimates 
of the standard deviation of $<q>$, not to be confused with the overall 
standard deviation $\sqrt{Var q}$ which is larger and is much more lengthy 
to compute.}
\label{fig_overlap}
\end{figure}

We denote the averages of $<q>$ and $<q^2>$ over the random strings
$X$ and $Y$ simply by $E(q)$ and $E(q^2)$. 
Figure (\ref{fig_overlap}) presents the results of a Monte Carlo 
computation of $E(q)$ and $Var(q)=E(q^2)-(Eq)^2$ in the case $S=2$. 
This figure shows that $E(q)$ has a nearly $1/\sqrt{N}$ convergence 
to a limit value $q_S$ as $N\to \infty$. 
Not surprisingly in view of the fact that ${\cal N}_{LCS}$ grows 
exponentially with $N$, we find that $q_S<1$. Estimates of 
$q_S$ based on a $1/\sqrt{N}$ extrapolation of our finite size results
are given in table (\ref{tab_alpha_q}).

It is also seen on figure (\ref{fig_overlap}) that the variance of the 
overlap decreases with $N$ roughly as $1/N$. 
Hence we conclude that the overlap $q(\sigma_1,\sigma_2)$ of two randomly 
chosen LCSs happens to be self-averaging, i.e. $q(\sigma_1,\sigma_2)$ 
becomes non random (and equal to $q_S$) in the limit $N\to \infty$. 
This is in fact not surprising: the space of LCSs of two random 
strings is not very far from having a product structure and the  
quantity $(1-q(\sigma_1,\sigma_2))$ is a kind of (normalized) Hamming 
distance on this space.
In the conventional wisdom of statistical mechanics, we would say that, 
although there is some pathology in this system from a 
physical point of view (it does not satisfy ``Nernst's principle''), 
it presents no replica symmetry breaking. 

\begin{table} 
\caption{The exponential growth factor of the number of LCSs
of two random strings and the average overlap between two LCSs.} 
\label{tab_alpha_q}
\begin{center}
\begin{tabular}{|c|cc|c|}
\hline
$S$ & $\alpha_S$ & $\beta_S$ & $q_S$\\
\hline
2 & 0.2458(8) & 0.232(2) & 0.6753(8)\\ 
3 & 0.2302(4) & 0.171(1) & 0.6782(8)\\ 
4 & 0.2086(3) & 0.145(2) & 0.6851(7)\\ 
5 & 0.1903(2) & 0.125(2) & 0.6921(7)\\ 
10 & 0.1365(2) & 0.0885(1) & 0.7138(10)\\ 
15 & 0.1100(1) & 0.0711(1) & 0.7264(8)\\
\hline
\end{tabular}
\end{center}
\end{table}

\begin{figure} 
\begin{center}
\resizebox{0.45\textwidth}{!}{\includegraphics{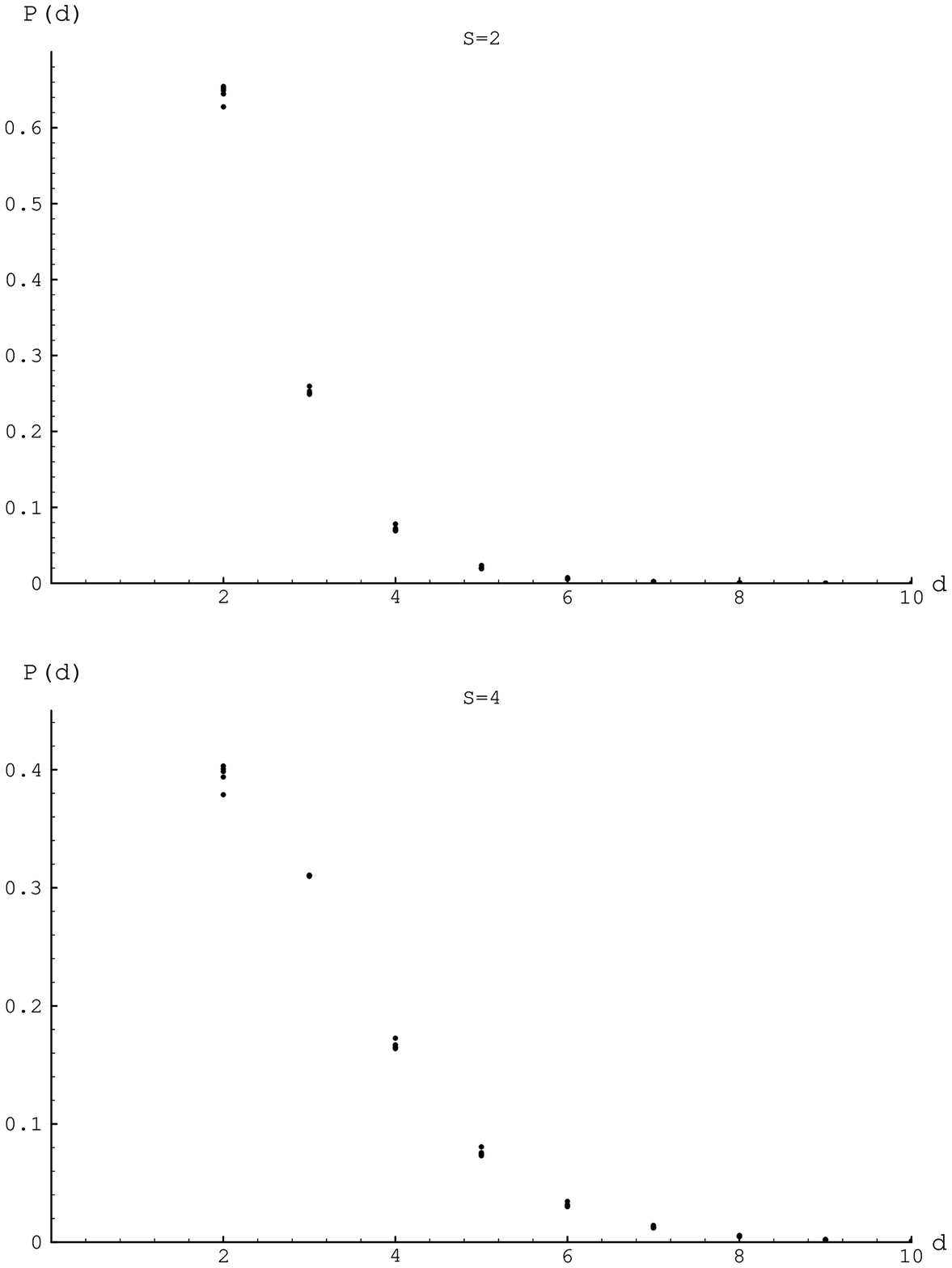}}
\end{center}
\begin{center}
\resizebox{0.45\textwidth}{!}{\includegraphics{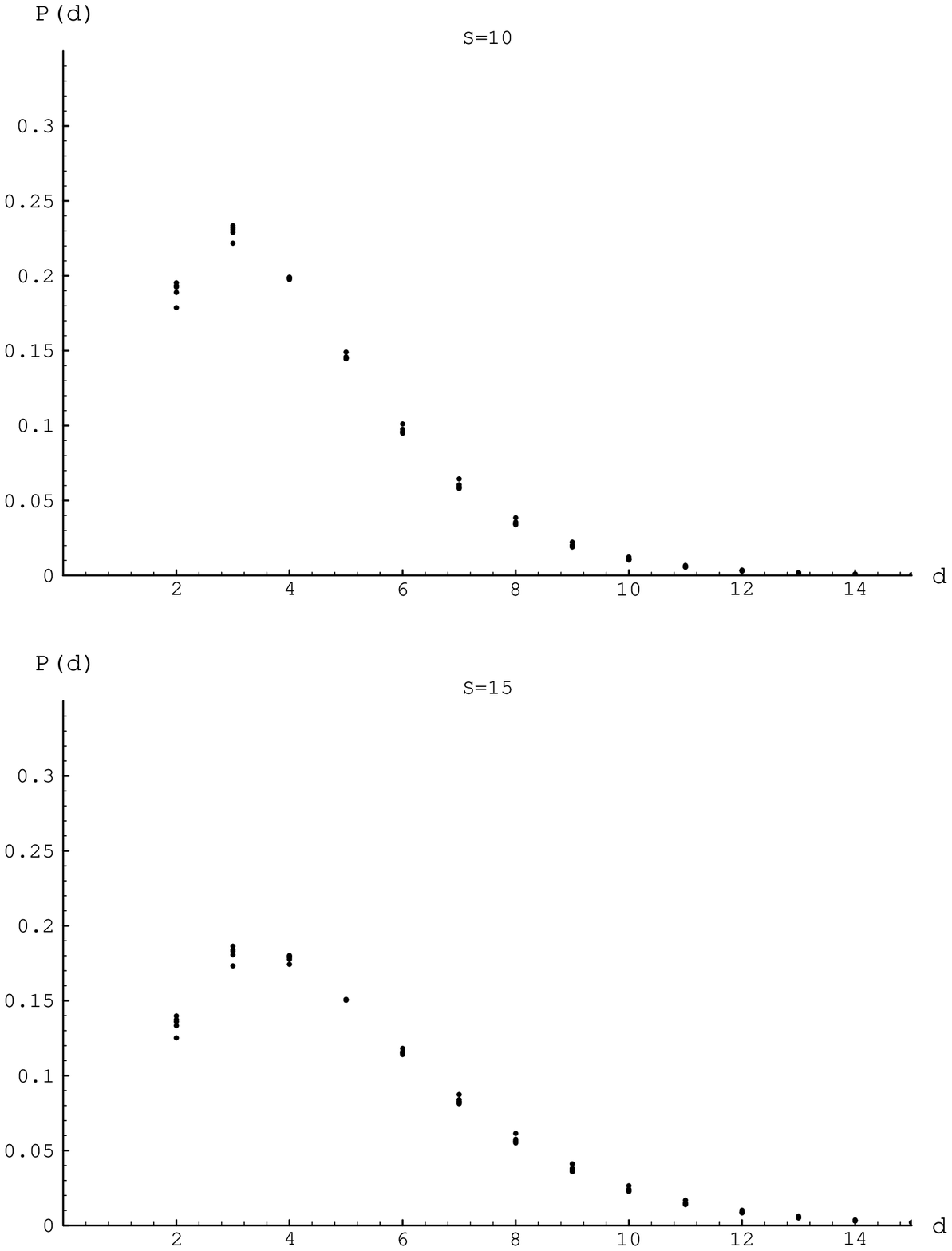}}
\end{center}
\caption{The distribution ${\cal P}_S(d)$ of the distance between 
two successive LCS matches, for $S=2,4,10,15$ 
(averages over $10^4$ random strings in each case). 
Each figure show results for different values of $100\le N\le 1500$ 
superposed in order to visualize the collapse toward a limit value.}
\label{fig_plcs}
\end{figure}

\begin{figure} 
\begin{center}
\resizebox{0.50\textwidth}{!}{\includegraphics{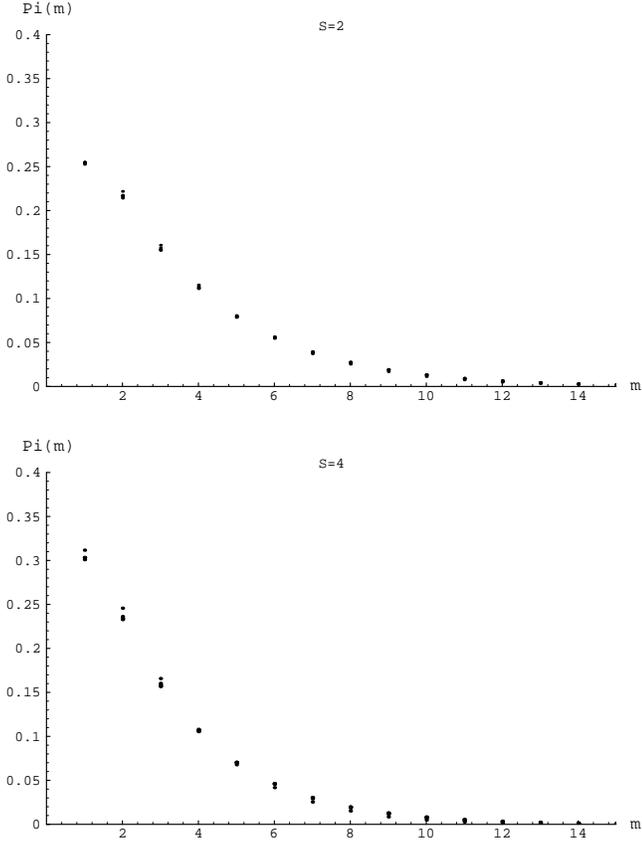}}
\end{center}
\caption{The distribution $\Pi_S(m)$ of the number of LCS matches of a 
given rank for $S=2$ and $S=15$ (averages over $10^4$ random strings in
each cases). }
\label{fig_prank}
\end{figure}

We also considered quantities which are of interest to 
describe the ``shape'' of the LCS graph.
Two such quantities are the distribution of the distance 
between two successive matches of a LCS, and 
the distribution of the number of LCS matches of a given rank. 
More precisely, we let ${\cal P}(d,X,Y)$ be 
the empirical distribution, over the set of LCSs of $X$ and $Y$,
of the distance between two successive LCS matches:
\begin{equation}
{\cal P}(d,X,Y)={1\over L-1} \sum_{k=1}^{L-1} \sum_{Q\in M_{LCS}(k)}
{{\cal N}_{LCS}(Q,d)\over {\cal N}_{LCS}}. 
\end{equation}
Here ${\cal N}_{LCS}(Q,d)$ is the number of LCS $\sigma=(Q_1,...,Q_L)$ of 
$X$ and $Y$ such that $Q_k=Q$ for some $k<L$, and $|Q_{k+1}-Q_k|=d$ 
(the distance between two points is taken to 
be Manhattan distance $|(i_1j_1)-(i_2j_2)|=|i_1-i_2|+|j_1-j_2|$).
We define ${\cal P}_S(d,N)$ as the average of ${\cal P}(d,X,Y)$ over
random $S$-ary strings of size $N$.
Also we let $\Pi(m,X,Y)$ be the empirical distribution of the 
cardinality of $M_{LCS}(k)$ over $1\le k\le L$, i.e.
\begin{equation}
\Pi(m,X,Y)={1\over L} \sum_{k=1}^L \delta(l_k,m),
\end{equation}
$l_k$ being the number of LCS matches of rank $k$, and we let 
$\Pi_S(m,N)$ be the average of $\Pi(m,X,Y)$ over $X$ and $Y$.
It is natural to expect that ${\cal P}_S(d,N)$ has a limit 
${\cal P}_S(d)$ as $N\to \infty$. 
It is not so obvious that the same holds for $\Pi_S(m,N)$. 
We found numerically that both ${\cal P}_S(d,N)$ and $\Pi_S(l,N)$ approach
well defined distributions as $N$ grows. Figure (\ref{fig_plcs}) 
reproduces graphically ${\cal P}_S(d)$ for $S=2,4,10$ and $15$. As $S$ 
increases the maximum of ${\cal P}_S(d)$ becomes more and more pronounced 
and is displaced to the right, as is expected from the relation
$\sum_{d\ge0}d{\cal P}_S(d)=2/\gamma_S$.
The asymptotic shape of $\Pi_S(m)$ appears to depend much less 
drastically on $S$ so we give only the results obtained for 
$S=2$ and $S=4$ (figure (\ref{fig_prank})). Numerically it is found that the 
typical number of LCS matches of a given rank remains bounded as
$N$ grows.

\begin{figure} 
\begin{center}
\resizebox{0.50\textwidth}{!}{\includegraphics{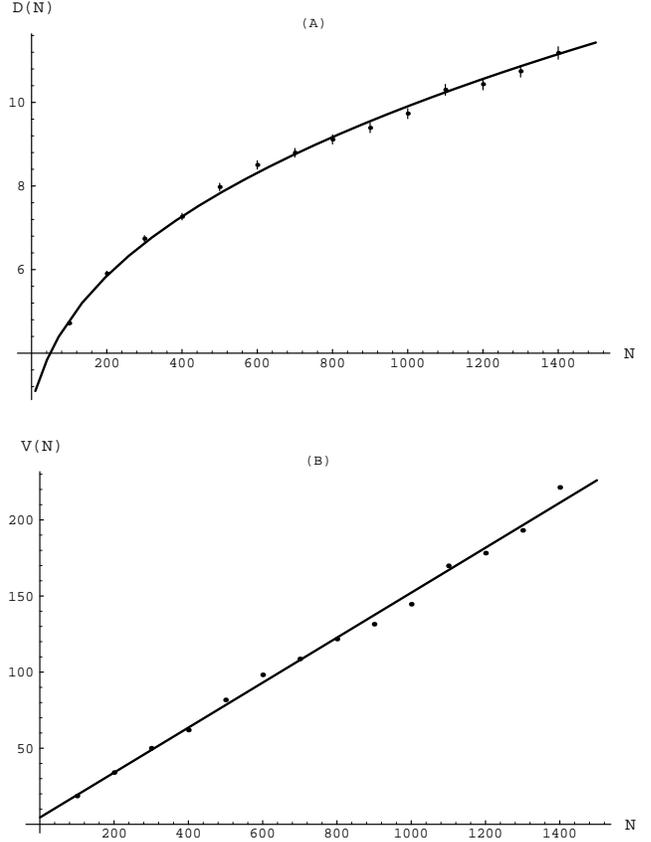}}
\end{center}
\caption{Behaviour of (A) the mean $D_S(N)$ and (B) the variance $V_S(N)$ 
of the width of the LCS graph (averages over $10^4$ random $15$-ary 
strings).}
\label{fig_drank}
\end{figure}

This contrasts with the behaviour of the {\it diameter} of the sets 
$M_{LCS}(k)$ (in the Manhattan distance). This behaviour is shown 
in figure Figure (\ref{fig_drank}), where are plotted the quantities 
$D_S(N)$ and $V_S(N)$, defined to be respectively the mean and 
variance over random $S$-ary strings of size $N$ of 
\begin{equation}
D_S(X,Y)={1\over L} \sum_{k=1}^L diam(M_{LCS}(k)).
\end{equation}
Clearly $D_S(N)$ appears to grow with $N$. 
In fact from heuristic scaling arguments, we expect $D_S(N)$ 
to be of the same order as the finite size corrections to the linear 
scaling of $E(L_N)$. If we are confident in (\ref{fsize_EL}), 
this means that $D_S(N)$ should grow asymptotically as $\sqrt{N}/\ln{N}$. 
Fortunately this is what we find from a $\chi^2$ analysis: 
The solid curve in figure (\ref{fig_drank})(A) is a best fit of 
our estimates to a function of the form $C_1 + C_2 \sqrt{N}/\ln{N}$. 
The corresponding $\chi^2$ value is $12,74$ for a number of degrees 
of freedom of $13$. For comparison, the $\chi^2$ value achieved from a 
best fit to $C_1+C_2\sqrt{N}$ is of $37.7$, which is much too large. 
This numerical test provides another support to the reliability of 
(\ref{fsize_EL}). Note however that the fluctuations of $D_S(X,Y)$ are 
far from negligible, as the variance of $D_S(X,Y)$ shows a near linear 
growth.

\begin{figure} 
\begin{center}
\resizebox{0.45\textwidth}{!}{\includegraphics{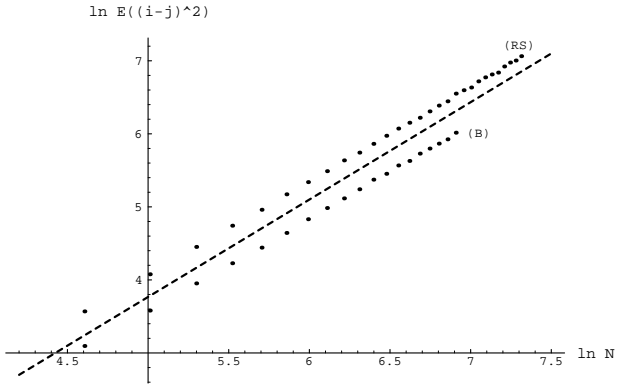}}
\end{center}
\caption{Scaling of the average ``displacement'' $E((i-j)^2)$ along the 
LCS graph, for the Random String model (RS) ($100\le N \le 1500$) and for 
the Bernoulli Matching model (B) ($100\le N\le 1000$). Averages are over $10^4$ 
instances in each case, with $S=2$. The $N^{4/3}$ scaling is visualized by the 
dashed line of slope $4/3$.}
\label{fig_disp}
\end{figure}

The asymptotic distributions ${\cal P}_S(d)$ and $\Pi_S(m)$ provide useful
informations on the local properties of the LCS graph, but they tell nothing 
about the universality class of the LCS problem.
Results for the mean square ``displacement'' $i-j$ along the LCS graph 
are presented in figure (\ref{fig_disp}). 
One way to measure this quantity would be to generate a given LCS
in a sequential way and to perform averages along this LCS 
\cite{DrasdoHwaLassig98_condmat}. 
Since we are able to perform exact averages over the set of LCSs, 
we use here a more ``static'' definition: For a given instance, 
the mean square displacement along a LCS chosen at random is given by
\begin{equation}
<(i-j)^2>={1\over L}\sum_{k=1}^L \sum_{Q\in M_{LCS}(k)} P_1(Q)(i-j)^2 
\end{equation}
where $P_1(Q)$ is defined as before and we have set $Q=(ij)$.
We then estimate $E((i-j)^2)$ as an average over a large number of random
strings of $<(i-j)^2>$ computed for each instance. The price to pay for 
exact computations is mainly a limitation on the size $N$ of our instances. 
It is seen on figure (\ref{fig_disp}) however that the scaling 
behaviour $E((i-j)^2)\approx N^{4/3}$ is reached rather fast, {\it both}
for the Bernoulli matching model and for the Random String model.
We cannot exclude the possibility of a crossover at $N\gg 1500$ 
for the Random String model. But then the asymptotic 
scaling regime of $E((i-j))^2$ would be attained at much larger 
values of $N$ than for $Var(L_N)$, which seems unlikely.

\section{Concluding remarks.}
\label{Conclusion}

This article has been devoted to the presentation of a thorough 
investigation of the LCS Problem by means of numerical simulations.
One of our main findings is that the finite size behaviour 
of the average LCS length $E(L_N)$ is very well reproduced 
by (\ref{fsize_EL}). This form provides a numerically 
trustworthy method of extrapolation, from which we have improved 
significantly the precision on previous estimates of the
limit ratio $\gamma_S$. 
It is very difficult at present to find any theoretical 
insight which could justify (\ref{fsize_EL}). Even improving on 
Alexander's rate result seems very difficult. 
It could be useful in this respect to have a better 
understanding on the effects of boundary conditions in these 
kind of problems. 

We also studied a related model where the two 
strings are replaced by a matrix of i.i.d. Bernoulli variables 
indicating the locations of the matches. We obtained a simple 
analytic expression (\ref{gammaBr_expression}) for the 
``passage time'' function $\gamma^B_S(r)$ of this Bernoulli 
Matching model. This expression compares very well with our numerical 
results, and it provides also an excellent approximation for the 
Random String model. As this approximation becomes more and more 
accurate as $S$ becomes large, a natural question is then whether one 
could evaluate some corrections induced by the correlations among 
matched points in the Random String model.
 
A further interesting question concerns the applicability of 
the cavity-like method used to derive ({\ref{gammaBr_expression}). 
What makes this method work for the Bernoulli Matching model is that 
a remarquable decorrelation property holds  in this percolation 
problem \cite{Boutet98_PRL}.
It would be interesting to find other percolation 
problems where such a decorrelation property occurs. 
This would provide simple means to obtain analytical 
information on the passage time constants of such models. 

We finally investigated average properties of the set of 
solutions, and the ``universality class'' of the LCS problem. 
We were rather surprised to find that the number of common 
subsequences of maximal size of two typical random strings 
grows exponentially with the size of the strings.
It follows that two (randomly) given LCSs are to a large extent 
distinct, as confirmed by the study of their typical overlap.  
We also found that the long ranged correlations 
in the Random String model appear to be relevant to the universality
class of the model, as is seen from the large $N$ behaviour of
$Var(L_N)$. One may wonder why this has not been observed in 
Needleman-Wunsch sequence alignment \cite{HwaLassig96_PRL}. A plausible 
reason (pointed out in \cite{HwaLassig96_PRL}) is that introducing a gap 
penalty in the model results in binding more tightly the optimal paths to the first bissector. This should reduce the effect of correlations and extend the ``small $N$'' scaling regime to larger values of $N$. In particular for biological purposes only the small $N$ regime is likely to be relevant. An exciting issue is the possible occurence of a phase transition in the gap
parameter of Needleman-Wunsch alignment. 

Another interesting question is whether a proliferation
of solutions is specific to random sequences and subsequences,
or if such phenomenon is of relevance to other percolation 
situations. As already said, the smallness of the variance of 
$L_N$ is probably related to the large number of LCSs of two random
sequences. Smallness of the variance of the passage time from 
$(0,0)$ to $(0,N)$ is also observed in usual first passage percolation 
on ${\bf Z}^2$. In fact these models (a famous example of which is the 
Eden model) are known to fall into the universality class of directed 
polymers in random media. One may expect to find in these models a large 
number of quasi optimal paths with typical overlaps smaller than one.

\section*{Aknowledgements}
J. BdM. wish to thank professor P. Blanchard for his
kindness and hospitality at the BiBoS center of Bielefeld,
and for several stimulating discussions. He is very grateful 
to D. Gandolfo who suggested the use of a parallel 
machine and initiated him to parallel computing, and to 
O.C. Martin for his useful comments on the paper.
It is a pleasure to thank M. Steele for useful communications. 
Thanks go also to one of the referees for pointing to the author 
reference \cite{HwaLassig96_PRL}. This work has been supported 
by the EU-TMR project ``Stochastic Analysis and its Applications''.

\bibliographystyle{perrot}
\bibliography{lcs_epjb}

\end{document}